\documentclass[prd,preprint,superscriptaddress,preprintnumbers,eqsecnum,showpacs,nofootinbib,nobibnotes]{revtex4}
\usepackage{amsfonts,amsmath,amssymb,bm,natbib}
\usepackage{graphicx} 

\newcommand{\be}{\begin{equation}}
\newcommand{\bea}{\begin{eqnarray}}
\newcommand{\ee}{\end{equation}}
\newcommand{\eea}{\end{eqnarray}}
\def\chic#1{{\scriptscriptstyle #1}}

\def\chic#1{{\scriptscriptstyle #1}}

\def\s#1{{\scriptscriptstyle #1}}

\def\noeq#1{(\ref{#1})}
\def\1eq#1{Eq.~(\ref{#1})}

\def\2eqs#1#2{Eqs.~(\ref{#1}) and~(\ref{#2})}
\def\3eqs#1#2#3{Eqs.~(\ref{#1}),~(\ref{#2}) and~(\ref{#3})}

\def\fig#1{Fig.~\ref{#1}}

\def\kint{\int_k\!}
\def\diff#1{{\rm d}^#1}

\def\ie{{\it i.e.}, }

\def\cd{\!\cdot\!}

\def\uE#1{{#1}^\s{\rm{E}}}
\def\dE#1{{#1}_\s{\rm{E}}}

\def\Gz{\Gamma^{abcd(0)}_{\mu\nu\rho\sigma}}
\def\G{G^{abcd}_{\mu\nu\rho\sigma}}
\def\T{T^{abcd}_{\mu\nu\rho\sigma}}
\def\R{R_{\mu\nu\rho\sigma}}

\def\Tra{{\rm Tr}_{\s{\rm{adj}}}}

\begin{document}
\title{Nonperturbative study of the four gluon vertex}

\author{D. Binosi}
%\email{binosi@ectstar.eu}
\author{D. Iba\~nez}
\affiliation{European Centre for Theoretical Studies in Nuclear
Physics and Related Areas (ECT*) and Fondazione Bruno Kessler, \\Villa Tambosi, Strada delle
Tabarelle 286, 
I-38123 Villazzano (TN)  Italy}

\author{J. Papavassiliou}
%\email{Joannis.Papavassiliou@uv.es}
\affiliation{\mbox{Department of Theoretical Physics and IFIC, 
University of Valencia and CSIC},
E-46100, Valencia, Spain}

\begin{abstract}
 
In this paper we study the nonperturbative structure of the SU(3) four-gluon vertex 
in  the Landau gauge, concentrating on contributions 
quadratic  in the metric.  
We employ an 
approximation scheme where ``one-loop''  diagrams are computed using  
fully  dressed gluon and ghost propagators, and tree-level vertices. 
When a suitable  kinematical configuration depending  on a
single  momentum  scale $p$  is  chosen,  only  two structures
emerge: the tree-level four-gluon vertex, and a tensor orthogonal
to it. 
A detailed numerical analysis reveals that the form factor associated with this 
latter tensor displays 
a change of sign (zero-crossing) in the deep infrared, 
and finally diverges logarithmically.  
The origin  of this characteristic behavior is  proven to be entirely
due to the masslessness of the ghost propagators forming the corresponding 
ghost-loop diagram, in close analogy to a similar effect 
established for the three-gluon vertex. 
However, in the case at hand, and under the approximations employed, 
this particular  divergence does not 
affect the form factor proportional to the tree-level tensor,  
which remains finite in the entire range of  momenta, 
and deviates moderately from its naive tree-level value.
It turns out that the kinematic configuration chosen is
ideal for carrying out lattice simulations, because it eliminates  
from the connected Green's   function all one-particle  reducible
contributions, projecting out the genuine one-particle irreducible 
vertex. 
Motivated by  this possibility,  we discuss  in
detail  how a  hypothetical  lattice measurement  of  this quantity  would
compare to the results presented here, and the potential interference  
from an additional tensorial structure, allowed by 
Bose symmetry, but not encountered within our scheme.

\end{abstract}

\pacs{
12.38.Aw,  % General properties of QCD (dynamics, confinement, etc)
12.38.Lg, % Other nonperturbative calculations
14.70.Dj %Gluons
}

\maketitle

\section{Introduction}

Of all elementary vertices that appear in the QCD Lagrangian, the four-gluon vertex is the most poorly understood. 
From the point of view of continuum studies, this fact may be regarded as 
a consequence of the enormous proliferation of allowed tensorial structures, generated by the 
presence of four color and four Lorentz indices. This difficulty, in turn, complicates considerably the extraction of reliable 
nonperturbative information from the corresponding Schwinger-Dyson equation (SDE). 
In addition, even gauge-technique inspired Ans\"atze~\cite{Salam:1963sa,Salam:1964zk,Delbourgo:1977jc,Delbourgo:1977hq} 
are extremely difficult to implement, 
due to the complicated structure of the Slavnov-Taylor identity that this vertex satisfies in the 
linear covariant ($R_{\xi}$) gauges (see, e.g.~\cite{Binosi:2009qm}).
Thus, the analytic studies dedicated to this vertex are very scarce, furnishing information only at the level 
of one-loop perturbation theory~\cite{Pascual:1980yu,Gracey:2014ola}, or involving generic constructions in the 
context of the pinch technique~\cite{Papavassiliou:1992ia}, or privileged 
quantization schemes, such as the background field method~\cite{Hashimoto:1994ct,Ahmadiniaz:2013rla}.

From the point of view of lattice simulations, 
the situation is simpler, in the sense that, to the best of our knowledge,  no simulations of the four-gluon vertex 
have been performed, for any kinematic configuration. This is to be contrasted with the corresponding status of all other vertices, 
namely the quark-gluon, the ghost-gluon, and three-gluon vertex, 
which have been studied on the lattice, at least for some special choices of their momenta~\cite{Skullerud:2002ge,Skullerud:2003qu,Cucchieri:2004sq,Sternbeck:2006rd,Cucchieri:2006tf,Cucchieri:2008qm}.

In the present work, we carry out a preliminary nonperturbative study of 
the  one-particle
irreducible (1-PI) part of 
the four-gluon vertex, denoted by ${\Gamma}^{abcd}_{\mu\nu\rho\sigma}$, 
motivated by recent developments in our understanding of the QCD nonperturbative dynamics of the two- and three-point sectors in the Landau gauge.
Specifically, a precise nonperturbative connection between the masslessness of the ghost, 
the detailed shape of the gluon propagator in the deep infrared (IR), 
and the IR divergences observed in certain kinematic limits of the three-gluon vertex, has been put forth in~\cite{Aguilar:2013vaa} (see also~\cite{Blum:2014gna,Eichmann:2014xya} for related contributions). 
This detailed study led to the conjecture 
that any purely gluonic $n$-point function 
will display the same kind of behavior, given that ghost loops\footnote{We refer to  ghost loops that exist already at the one-loop level. 
Ghost loops nested within gluon loops do not produce this particular effect, because the 
additional integrations over virtual momenta soften the IR divergence.} 
appear in all of them (and, hence, the associated IR logarithmic divergence in $d=4$).
Clearly, the confirmation of this expectation at the level of the 
four-gluon vertex would put our understanding of this specific IR effect on rather solid ground. 
In particular, it would be important to establish, even within an approximate scheme, 
the type of tensorial structures that will be associated 
with this particular divergence.

In order to simplify the calculation 
as much as possible without compromising its main objective, 
we have chosen a particularly simple configuration of the external momenta, in which a single momentum scale ($p$) appears, and the flow in the four legs is chosen to be $(p,p,p,-3p)$; this has the advantage of giving rise to loop integrals that are symmetric under the crossing of external legs thus reducing the amount of diagrams one needs to evaluate. 
We hasten to emphasize that the aforementioned momentum configuration has been first considered in~\cite{Kellermann:2008iw}, in the context of the so-called ``scaling'' solutions~\cite{Alkofer:2000wg}. Instead, our analysis will be carried out using an IR finite gluon propagator $\Delta$ and ghost dressing function $F$, in conformity with the results obtained from a plethora of large-volume lattice simulations~\cite{Cucchieri:2007md,Sternbeck:2007ug,Bowman:2007du,Bogolubsky:2009dc,Oliveira:2009eh,Cucchieri:2010xr,Ayala:2012pb}, as well as a variety of analytic approaches~\cite{Alkofer:2000wg,Szczepaniak:2001rg,Maris:2003vk,Szczepaniak:2003ve,Aguilar:2004sw,Fischer:2006ub,Kondo:2006ih,Braun:2007bx,Binosi:2007pi,Epple:2007ut,Boucaud:2008ky,Binosi:2008qk,Aguilar:2008xm,Fischer:2008uz,Szczepaniak:2010fe,Watson:2010cn,RodriguezQuintero:2010wy,Campagnari:2010wc,Pennington:2011xs,Watson:2011kv,Kondo:2011ab,Aguilar:2011ux,Binosi:2012sj}. Specifically, we will consider a simplified version of the 
so-called ``one-loop dressed'' approximation, where one computes 
the one-loop diagrams with 
fully dressed gluon and ghost propagators, but with tree-level (undressed) 
vertices (the case with  dressed ghost-vertices only is also presented).

Notice that this approach, although SDE-inspired, differs significantly from a typical SDE study,  mainly because it does not involve the solution of an integral equation for the unknown form factors; instead, the form factors are simply extracted from the dressed diagrams mentioned above. In that sense,  it may be thought of as a ``lowest order'' SDE approximation, where one simply substitutes tree-level values for all vertex form factors appearing inside diagrams. This particular method (and variations thereof) has been employed in the context of other vertices, furnishing results that compare favorably with the lattice~\cite{Aguilar:2013vaa,Aguilar:2013xqa,Aguilar:2014lha}; of course, its effectiveness  can only be justified {\it a-posteriori} (\ie comparing with the lattice), given that there is no rigorous way of estimating the errors introduced by the omitted terms.

If  one concentrates  on the nonperturbative  contributions that are quadratic in the metric, in the case of SU(3) only two independent tensorial structures  emerge: the  one associated with  the tree-level four-gluon vertex (indicated by $\Gz$),  and a second one (denoted with $\G$) which is totally symmetric in both Lorentz as well color indices
(and therefore orthogonal  in both spaces to the  tree-level term). It turns out that the aforementioned divergences are entirely proportional to  this latter tensorial structure,  with no contribution to the tree-level  tensor $\Gamma^{(0)}$. Therefore, one finds  that within the one-loop dressed approximation  we employ, $G$ will carry  all the IR divergences, whilst $\Gamma^{(0)}$ contains all the ultraviolet (UV) divergences,  
as required by the renormalizability of the theory.
These findings clearly deviate from the patterns 
observed in the case of the three-gluon vertex, where the form factors proportional to the tree-level vertex, 
in addition to containing the UV divergences, 
were also affected by this particular IR divergence (displaying the associated ``zero crossing'').
In addition,  the deviation of the form factor associated to the tree-level tensor $\Gamma^{(0)}$ from 1, namely its tree-level value, is relatively modest. In particular, when the ingredients used in its calculation are renormalized at $\mu=4.3$ GeV,
its highest value, located at about 500 MeV, is 1.5.  

The results obtained  are further discussed in the specialized context of a
possible future lattice  simulation  of  the {\it connected}  part of this vertex,
to be denoted by ${\cal   C}^{abcd}_{\mu\nu\rho\sigma}$. 
It turns out that the momentum
configuration  $(p,p,p,-3p)$ eliminates   all  contributions  to   ${\cal  C}$  from
one-particle reducible (1-PR)  graphs, thus isolating only ${\Gamma}^{abcd}_{\mu\nu\rho\sigma}$, 
without any ``contamination'' from lower-order Green's functions.
In addition, an analysis based  on Bose symmetry arguments alone, reveals
that a  third tensor structure,  denoted by $X'^{abcd}_{\mu\nu\rho\sigma}$, is in
principle  allowed;  evidently, the form factor
associated with this  tensor vanishes  within the  one-loop  dressed  approximation
that we employ. It is likely, however, 
that this particular property will not 
persist in a complete nonperturbative computation, as the  one provided by 
lattice simulations. Therefore, under the assumption that such  a structure might eventually emerge,  
we describe how to express the complete set  of form  factors characterizing ${\Gamma}^{abcd}_{\mu\nu\rho\sigma}$ in
terms  of  the  standard  lattice  ratios $R$,  used in 
studies of the three-gluon vertex~\cite{Cucchieri:2006tf,Cucchieri:2008qm}.

The article is organized as follows. In Sect.~\ref{sec:general} we introduce our notation, review the  
relevant tensor decomposition, and recall some identities particular to the SU(3) 
gauge group. Next, in Sect.~\ref{sec:zeroext} we carry out the calculation of the one-loop dressed diagrams 
in the simplified setting where all the external momenta are set to zero. This will prove to be a very 
useful exercise, as it will allow to determine the tensorial structures that appear, and in particular 
establish that the divergent part coming from ghost loops is entirely proportional to the $\G$ tensor alone. 
Then, in Sect.~\ref{sec:nonzeroext} we carry out the calculation in the $(p,p,p,-3p)$ momentum configuration. 
After manipulating all diagrams analytically (Sect.~\ref{sec:analytic}), we evaluate  
numerically all the contributions obtained, using (quenched) lattice 
results as input for the gluon and ghost two-point sectors (Sect.~\ref{sec:numeric}). Finally, in Sect.~\ref{sec:lattice} 
we show how our results can be related to quantities customarily studied on the lattice. 
Specifically, we prove that the special momentum configuration chosen for our study 
has the property of isolating the 1-PI  contribution to the connected four-gluon Green's function. Then, assuming the most general tensor decomposition of this vertex in terms of tensors allowed by Bose symmetry, we show what would be the best choice of the ratios $R$.
The paper ends with Sect.~\ref{concl}, where we draw our conclusions, 
and two Appendices. In the first, we carry out a general analysis of the tensor structures (quadratic in the metric) that are allowed by Bose symmetry,
paying particular attention to the case $(p,p,p,-3p)$.
Finally, Appendix~\ref{sif} collects some lengthy expressions appearing in our analytical calculations.   

\section{\label{sec:general}Generalities on the four-gluon vertex}

\begin{figure}[!t]
\includegraphics[scale=0.65]{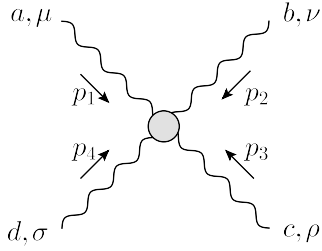}
\caption{\label{fig:4g-tree-level}The 1-PI 4-gluon vertex; we display the momenta [note that $p_4= -(p_1+p_2+p_3)$],  
together with the Lorentz and color indices.}
\end{figure}

The 1-PI four-gluon vertex will be denoted by the expression (all momenta entering)
\begin{equation}
\Gamma_{A^a_\mu A^b_\nu A^c_\rho A^d_\sigma}(p_1,p_2,p_3,p_4)=-ig^2\Gamma_{\mu\nu\rho\sigma}^{abcd}(p_1,p_2,p_3,p_4).
\label{4g-general}
\end{equation}
At tree-level one has
\begin{align}
\Gamma_{\mu\nu\rho\sigma}^{abcd(0)}&=f^{adr}f^{cbr}(g_{\mu\rho}g_{\nu\sigma}-g_{\mu\nu}g_{\rho\sigma})+f^{abr}f^{rdc}(g_{\mu\sigma}g_{\nu\rho}-g_{\mu\rho}g_{\nu\sigma})\nonumber \\
&+f^{acr}f^{dbr}(g_{\mu\sigma}g_{\nu\rho}-g_{\mu\nu}g_{\rho\sigma}),
\label{gamma0}
\end{align}
where $f^{abc}$ are the real and totally antisymmetric SU(N) structure constants, satisfying the normalization condition
\begin{equation}
f^{ars}f^{brs}=N\delta_{ab}\,, 
\end{equation}
so that the generators of the adjoint representation are given by 
\be
(T_a)_{bc} = -i f_{abc}.
\label{Tadj}
\ee
In \fig{fig:4g-tree-level} we show the conventions of momenta and  
Lorentz/color indices used throughout this paper. 

Note that, due to Bose symmetry, $\Gamma_{\mu\nu\rho\sigma}^{abcd}(p_1,p_2,p_3,p_4)$ remains unchanged under the simultaneous
interchange of a set of its indices and momenta (e.g. $(a,\mu,p_1)\leftrightarrow (b,\nu,p_2)$, etc).
It is elementary to verify the validity of this symmetry for the tree-level vertex $\Gamma_{\mu\nu\rho\sigma}^{abcd(0)}$.

It is clear that the fully dressed $\Gamma_{\mu\nu\rho\sigma}^{abcd}$ is characterized, in general,  
by a vast proliferation of the tensorial structures (138 for general kinematics~\cite{Gracey:2014ola}); 
of course, as we will see, Bose symmetry imposes restrictions on the structure of the possible form factors composing $\Gamma_{\mu\nu\rho\sigma}^{abcd}(p_1,p_2,p_3,p_4)$.   

At the level of the rank-4 Minkowski tensors, the structures allowed are terms quadratic in the metric, 
linear in the metric and quadratic in the momenta, and quartic in momenta; schematically one has then the structures
\begin{equation}
gg;\qquad gpq;\qquad pqrs.
\end{equation}
At the level of the rank-4 color tensors the situation is considerably more complex, 
since, in addition to terms quadratic in $f$ or $\delta$,  the real and totally symmetric tensors $d_{abc}$ will also emerge. 
Thus, in principle one has 15 allowed structures of the schematic type
\begin{equation}
ff;\qquad dd;\qquad fd;\qquad \delta\delta.
\end{equation}
However, these tensors are related by a set  of 6 identities~\cite{Pascual:1980yu}, 
namely 
\begin{align}
&f_{abr}f_{cdr} = \frac{2}{N} \left[\delta_{ac} \delta_{bd} - \delta_{ad} \delta_{bc} \right] + d_{acr}d_{dbr} - d_{adr}d_{bcr},
\\
&f_{abr}d_{cdr} + f_{acr}d_{dbr} + f_{adr}d_{bcr} =0,
\label{ident}
\end{align}
and two independent permutation for each, 
a fact that reduces the number of required tensors down to 9.

Of course, due to practical limitations, one must restrict the present study to a considerably more reduced (but physically relevant) 
subset of the full Lorentz and color tensorial basis mentioned above. 
Specifically, as was done in~\cite{Pascual:1980yu}, we only consider 
terms quadratic in the metric tensor $g_{\mu\nu}$, namely terms proportional to $g_{\mu\nu} g_{\rho\sigma}$, $g_{\mu\rho} g_{\nu\sigma}$ , 
and $g_{\mu\sigma} g_{\nu\rho}$, neglecting terms quadratic and quartic in the momenta.
Thus, a priori, 
for a general SU(N) gauge group,  one has $9 \times 3=27$ possible combinations.  
Furthermore, we will directly specialize our analysis to the case $N=3$, where the additional identity 
\begin{equation}
\delta^{ab}\delta^{cd}+\delta^{ac}\delta^{bd}+\delta^{ad}\delta^{bc}=3[d_{abr}d_{cdr}+d_{acr}d_{bdr}+d_{adr}d_{bcr}]
\label{N3}
\end{equation}
can be used, thus reducing the number of tensorial combinations down to 24.

However, it turns out that, within the one-loop 
dressed approximation and the kinematical configuration that we will employ (see \fig{fig:4g-1loop-dressed} for the 
18 diagrams appearing in this case), the color tensors reduce finally to the 
two structures appearing in the conventional one-loop calculation of this vertex (for $N=3$), namely 
the tree-level tensor $\Gamma^{(0)}$ defined in~\1eq{gamma0}, and the totally symmetric (both in Minkowski and color space) tensor
\begin{equation}
\G=(\delta^{ab}\delta^{cd}+\delta^{ac}\delta^{bd}+\delta^{ad}\delta^{bc})\underbrace{(g_{\mu\nu}g_{\rho\sigma}+g_{\mu\rho}g_{\nu\sigma}+g_{\mu\sigma}g_{\nu\rho})}_{\R}.
\label{G-def}
\end{equation}
In particular, notice that since the two tensors are orthogonal in both spaces
\begin{align}
\Gamma^{abcd(0)}_{\mu\nu\rho\sigma}G^{mnrs}_{\mu\nu\rho\sigma}&=0;&
\Gamma^{abcd(0)}_{\mu\nu\rho\sigma}G^{abcd}_{\alpha\beta\gamma\delta}&=0,
\label{ortho}
\end{align}
the prefactors multiplying them can be unambiguously identified\footnote{As shown in Appendix~\ref{Bose}, 
Bose symmetry allows an additional tensor structure to appear; 
the consequences of this fact will be briefly addressed in~Sect.~\ref{sec:lattice}.}. 

Let us finally point out that, in SU(3), one has the additional useful formula
\begin{equation}
\Gz+\G=2X_{\mu\nu\rho\sigma}^{abcd},
\label{gagx}
\end{equation}
where we have defined the combination
\begin{align}
X_{\mu\nu\rho\sigma}^{abcd}&=\left(\delta^{ab}\delta^{cd}+\frac32 d_{abr}d_{cdr}\right)g_{\mu\nu}g_{\rho\sigma}+\left(\delta^{ac}\delta^{bd}+\frac32d_{acr}d_{bdr}\right)g_{\mu\rho}g_{\nu\sigma}\nonumber\\
&+\left(\delta^{ad}\delta^{bc}+\frac32d_{adr}d_{bcr}\right)g_{\mu\sigma}g_{\nu\rho}.
\end{align}

Our analysis of the four-gluon vertex will be carried out in the Landau gauge, where the study 
of the lower Green's functions (such as gluon and ghost propagator, 
ghost-gluon vertex and three-gluon vertex) has been traditionally carried out, both in the continuum as 
well as on the lattice. In this particular gauge the full gluon propagator takes the form
\be
i\Delta_{\mu\nu}(q)=-iP_{\mu\nu}(q)\Delta(q^2);\qquad 
P_{\mu\nu}(q)=g_{\mu\nu}-q_\mu q_\nu/q^2, 
\ee
while the ghost propagator, $D(q^2)$, and its dressing function, $F(q^2)$, are related by 
\be
D(q^2)= \frac{F(q^2)}{q^2}.
\label{ghostdress}
\ee
Evidently, both $\Delta(q^2)$ and $D(q^2)$ constitute crucial 
ingredients for the calculations of the four-gluon vertex that follows. 
It is therefore useful to briefly review some of 
their IR features that are most relevant to the present work. Specifically, 
both large-volume lattice simulations and a plethora of continuous nonperturbative 
studies, carried out both in SU(2) and in SU(3), converge to the conclusion 
that the function $\Delta(q^2)$
reaches a finite (nonvanishing) value in the IR.  
Moreover, the nonperturbative ghost propagator
remains ``massless'', and displays no IR enhancement, since its dressing function  $F(q^2)$ saturates 
in the deep IR to a finite value. 
As we will see in what follows, the aforementioned features have far reaching consequences for the 
IR behavior of the four-gluon vertex. Specifically, as happens with the tree-gluon vertex, 
the masslessness of the ghost-loops contributing to $\Gamma_{\mu\nu\rho\sigma}^{abcd}$ 
produces a logarithmic IR 
divergence. What is, however, qualitatively distinct compared to the three-gluon case, 
is that, at least within the 
approximation scheme that we employ,  
this particular divergence does not manifest itself in the part proportional to $\Gamma^{(0)}$, 
but rather in the orthogonal combination $G$.

\section{\label{sec:zeroext}Vanishing external momenta} 

In this section we consider the simplest possible kinematic case, 
where all the momenta of the external gluons are set to zero ($p_1=p_2=p_3=p_4=0$).

\subsection{\label{sec:calc} The calculation}
Since we do no consider the contribution of quark-loops (pure Yang-Mills theory), the only 
representation that appears in our problem is the adjoint, whose explicit realization is given in \1eq{Tadj}.

For the various integrals appearing in this calculation we will employ the standard text-book results  
\begin{align}
\int_k f(k^2)k_\mu k_\nu &=  \frac1{d}g_{\mu\nu} \int_k k^2 f(k^2) 
\nonumber\\
\int_k f(k^2) k_\mu k_\nu k_\rho k_\sigma &=  \frac1{d(d+2)}\R \int_k k^4 f(k^2)
\end{align}
where $\R$ has been defined in~\1eq{G-def}, 
and the integral measure 
is \mbox{$\int_k=\mu^\epsilon\int\!\diff{d}k/(2\pi)^d$}, 
with $d=4-\epsilon$ the space-time dimension\footnote{Notice that 
we set $d=4-\epsilon$ instead of $d=4+2\epsilon$ used in~\cite{Pascual:1980yu}.} and $\mu$ the 't Hooft mass.

\begin{figure}[!t]
\includegraphics[scale=0.775]{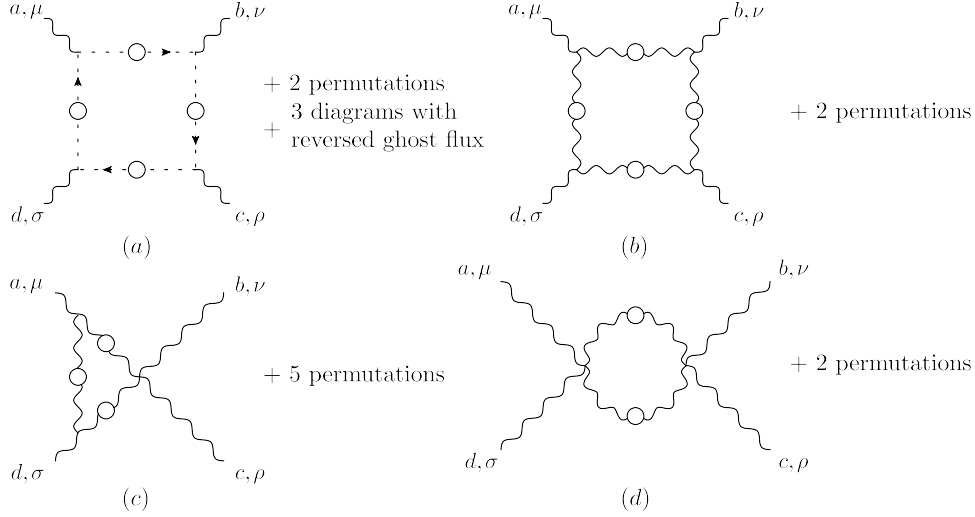}\caption{\label{fig:4g-1loop-dressed}The 18 diagrams contributing to the four-gluon vertex in the one-loop dressed approximation. The fishnet diagrams $(d)$ carry a statistical factor of $1/2$. Lorentz, color and momentum flow are as in~\fig{fig:4g-tree-level}.}
\end{figure}

There are two particular tensorial structures that appear in a natural way in the calculations of the graphs shown in~\fig{fig:4g-1loop-dressed}, namely  
\begin{align}
Q_{1\,\mu\nu\rho\sigma}^{abcd} &\equiv  \Tra[(T^aT^bT^cT^d)g_{\mu\rho}g_{\nu\sigma}+(T^aT^cT^bT^d)g_{\mu\nu}g_{\rho\sigma}
+(T^aT^bT^dT^c)g_{\mu\sigma}g_{\nu\rho}],
\nonumber\\
Q_{2\,\mu\nu\rho\sigma}^{abcd} &\equiv \Tra[(T^aT^bT^cT^d)+(T^aT^dT^bT^c)+(T^aT^cT^dT^b)]\R.
\label{bastens}
\end{align}
Then, using the relation\footnote{Note also the particular property $\Tra(T^aT^bT^cT^d) =\Tra(T^aT^dT^cT^b)$, which is a consequence of the antisymmetric nature of the $T_a$ in \1eq{Tadj}, and can be directly verified using \1eq{adjtr}}
\begin{equation}
\Tra (T^aT^bT^cT^d)=\delta_{ab}\delta_{cd}+\delta_{ad}\delta_{bc}+\frac34(d_{abr}d_{cdr}-d_{acr}d_{bdr}+d_{adr}d_{bcr}),
\label{adjtr}
\end{equation}
together with~\1eq{gagx}, it is straightforward to express these structures in terms of $\Gamma^{(0)}$ and $G$,    
\begin{align}
Q_{1\,\mu\nu\rho\sigma}^{abcd} & = -\frac12\Gz+\frac34\G,
& Q_{2\,\mu\nu\rho\sigma}^{abcd} & = \frac94\G.
\label{identfin}
\end{align}

Turning to the explicit calculation of the one-loop dressed diagrams of~\fig{fig:4g-1loop-dressed}, the (six) ghost boxes give the result
\begin{equation}
\sum_{i=1}^6(a_i)^{abcd}_{\mu\nu\rho\sigma}=-2g^2\,\Tra [(T^aT^bT^cT^d)+(T^aT^dT^bT^c)+(T^aT^cT^dT^b)]\int_kk_\mu k_\nu k_\rho k_\sigma D^4(k^2),
\end{equation}
which, with the aid of the formulas~\noeq{identfin} introduced above, may be written in the simple form 
\begin{align}
\sum_{i=1}^6(a_i)^{abcd}_{\mu\nu\rho\sigma}&=g^2\G A(0);&
A(0)&=-\frac{9}{2d(d+2)}\kint\frac{F^4(k^2)}{k^4}.
\label{ghostboxes}
\end{align}
Since the ghost dressing function $F$ is known to saturate in the IR, the integral above diverges logarithmically in the IR; 
however~\1eq{ghostboxes} shows that this divergence does not contribute 
to the structures proportional to the tree-level tensor $\Gamma^{(0)}$. 
Even though this result has been derived in a simplified setting, 
it will persist within the one-loop dressed approximation employed here.
Therefore, we arrive at the important conclusion that the IR divergent 
terms originating from the ghost loops would be completely missed, 
if one were to consider only the form factor proportional to the tree-level tensor~$\Gamma^{(0)}$.

We next consider the (three) gluon boxes; 
as the adjoint traces will be the same as those appearing in the ghost case above, 
we obtain that also the one-loop dressed gluon boxes do not contribute to the tree-level tensor structure. In particular, we get
\begin{align}
\sum_{i=1}^3(b_i)^{abcd}_{\mu\nu\rho\sigma}&=g^2 \G B(0);&
B(0)&= \frac{36(d-1)}{d(d+2)}\kint k^4\Delta^4(k^2).
\label{gluonboxes}
\end{align}
Notice that, unlike the case of the ghost boxes treated above,
the  integral appearing in~\1eq{gluonboxes} is convergent in the IR, because 
the gluon propagator reaches a finite value in that limit. 

We now turn to the (six) triangle diagrams. After some straightforward algebraic manipulations, one obtains
\be
\sum_{i=1}^6(c_i)^{abcd}_{\mu\nu\rho\sigma} = 8g^2 
\left [\frac{d-2}{d} Q_{1\,\mu\nu\rho\sigma}^{abcd} - \frac{1}{d(d+2)} Q_{2\,\mu\nu\rho\sigma}^{abcd}  \right] \kint k^2\Delta^3(k^2),
\ee
which, after using the identities~\noeq{identfin}, can be cast in the form 
\begin{equation}
\sum_{i=1}^6(c_i)^{abcd}_{\mu\nu\rho\sigma}=g^2 \Gz C_1(0)+g^2\G C_2(0),
\label{zero-3g}
\end{equation}
where
\begin{align}
C_1(0)&=-\frac 4d(d-2)\kint k^2\Delta^3(k^2);&
C_2(0)=-12(d^2-1)\kint k^2\Delta^3(k^2).
\end{align}

Finally, we are left with the (three) fishnet diagrams. One finds, similarly to what happens with the triangle diagrams,
\be
\sum_{i=1}^3(d_i)^{abcd}_{\mu\nu\rho\sigma} = g^2
\left [\frac{6(d-2)}{d}\Gz -(d-2) Q_{1\,\mu\nu\rho\sigma}^{abcd} + \frac{d^3-4d+2}{d(d+2)} Q_{2\,\mu\nu\rho\sigma}^{abcd}  \right] \kint\Delta^2(k^2).
\ee
The identities~\noeq{identfin} allow us to express the result in its final form, namely  
\begin{equation}
\sum_{i=1}^3(d_i)^{abcd}_{\mu\nu\rho\sigma}=g^2\Gz D_1(0)+g^2\G D_2(0),
\end{equation}
with
\begin{align}
D_1(0)&=\frac{(d-2)(d+12)}{2d}\kint\Delta^2(k^2);&
D_2(0)&=\frac{3(d^3-4d+3)}{2d(d+2)}\kint\Delta^2(k^2).
\end{align}
The results obtained are conveniently summarized in Table~\ref{tab:1lres}. 

\begin{table}[!t]
\begin{tabular}{llll}
\hline\hline
{\bf Diagrams}\hspace{1cm} & {\bf Integral}\hspace{3cm} & $\frac{1}{d} g^2{\Gz}$\hspace{3cm} & $\frac1{d(d+2)}g^2{\G}$ \\
\hline
$(a)$ & $\kint\frac{F^4(k^2)}{k^4}$ & 0 & $-\frac92$ \\
$(b)$ & $\kint k^4\Delta^4(k^2)$ & 0 & $36(d-1)$ \\
$(c)$ & $\kint k^2\Delta^3(k^2)$ & $- 4 (d-2)$ & $-12(d^2-1)$\\
$(d)$ & $\kint \Delta^2(k^2)$ & $\frac{1}{2}(d-2)(d+12)$ & $ \frac{3}{2}(d^3-4d+3)$ \\
\hline
\end{tabular}
\caption{\label{tab:1lres} Contributions of the various class of diagrams to the four-gluon vertex in the one-loop dressed approximation with all external momenta set to zero.}
\end{table}

\subsection{\label{sec:pert} Perturbative analysis}

At this point one may explore the qualitative behavior of the two contributions obtained above 
within a setting inspired by one-loop perturbation theory, but supplemented by 
a set of mass scales, which prevent the resulting expressions from diverging in the 
IR. Specifically, if one were to simply 
set $F(k^2)$ and $\Delta(k^2)$ to their strict perturbative values ($1$ and $1/k^2$, respectively)  
the four integrals appearing in the second column of Table~\ref{tab:1lres} reduce to a 
single integral, namely $\kint\,\frac{1}{k^4}$. At this point,  it is easy to verify that, when $d=4$, the total contribution proportional to $\G$ vanishes, given that the sum of the coefficients appearing on the fourth column adds up to zero.

However, given that the integral $\kint\,\frac{1}{k^4}$ is both IR and UV divergent, 
it is preferable to introduce a distinction between the two type of divergences. To accomplish this, 
we proceed as follows. Given that the (Euclidean) gluon propagator (in the Landau gauge) is  known to be finite in the 
IR (a feature that can be self-consistently explained through the dynamical generation of an effective gluon mass),
for the purposes of this simple calculation one may approximate $\Delta(k^2)$ by $1/(k^2+m^2)$.
This replacement makes the integrals $\kint k^4\Delta^4(k^2)$, 
$\kint k^2\Delta^3(k^2)$, and $\kint \Delta^2(k^2)$ of Table~\ref{tab:1lres} IR finite; 
of course, they still diverge logarithmically in the UV. Regarding the integral $\kint\frac{F^4(k^2)}{k^4}$,
it is known that the ghost remains nonperturbatively massless, a fact that leads to a genuine IR divergence; in order to control 
it, we will introduce an artificial mass scale, denoted by $\lambda^2$. Thus, the integral corresponding to       
$\kint\frac{F^4(k^2)}{k^4}$ will read~$\kint\frac{1}{(k^2+\lambda^2)^2}$. 

Let us emphasize at this point that even though at the formal level both  $m^2$ and $\lambda^2$ 
serve as IR regulators, there is a profound physical difference between the two: $m^2$ constitutes a simplified 
realization of a true physical phenomenon, namely the IR saturation of the gluon propagator, 
while  $\lambda^2$ is an artificial scale, introduced as a regulator of a quantity 
(the ghost propagator) that is genuinely massless. Consequently, in order to recover the physically relevant 
(albeit simplified) limits, $m^2$ will be kept at some fixed 
nonvanishing value, while $\lambda^2$ will be sent to zero.  

The above considerations motivate the introduction of a particular integral, namely 
\begin{align}
I(M^2) &\equiv \kint\frac{1}{(k^2+M^2)^2}
\nonumber\\
& =  \frac{i}{16\pi^2} \left[\left(\frac2{\epsilon}-\gamma\right) - \ln (M^2/\mu^2) + {\cal O}(\epsilon)\right]\,, 
\label{IM}
\end{align}
where $\mu$ is the 't Hooft mass, and $\gamma$ the Euler-Mascheroni constant. 
Evidently, depending on the case that one considers,  $M^2=m^2$ or $M^2= \lambda^2$. 

In particular, after the replacements mentioned above, the integrals in Table~\ref{tab:1lres} can be expressed in terms of $I(M^2)$ 
as follows  
\begin{align}
\kint\frac{F^4(k^2)}{k^4} & \to  I(\lambda^2);&
\kint \Delta^2(k^2) & \to  I(m^2);\nonumber\\
\kint k^4\Delta^4(k^2) & \to  I(m^2) + \cdots;&
\kint k^2\Delta^3(k^2) & \to  I(m^2) + \cdots,
\end{align}
where the ellipses in the last two expressions indicate linear combinations of the integrals\footnote{These latter integrals 
appear simply through the elementary algebraic manipulation $k^2 = (k^2+m^2) - m^2$ in the numerators, and the 
subsequent cancellation of some of the denominators.} 
$m^2\kint\frac{1}{(k^2+m^2)^3}$ or $m^4\kint\frac{1}{(k^2+m^2)^4}$, which are convergent both in the IR and the UV.

At this point one may add up the corresponding contributions in the third and fourth columns 
of Table~\ref{tab:1lres} and obtain, within this perturbative scheme, 
the coefficients multiplying $\Gamma^{(0)}$ and $G$, to be denoted by $V_{\Gamma^{(0)}}^{(1)}(0)$ 
and $V^{(1)}_{G}(0)$, respectively. Specifically, setting $d=4$ everywhere (but keeping $\epsilon \neq 0$ in $2/\epsilon$), 
introducing $\alpha_s \equiv g^2/4\pi$, factoring out a $(-i)$ to conform with the definition in \1eq{4g-general},
we find for the leading behavior 
\be
V_{\Gamma^{(0)}}^{(1)}(0) =  2 i g^2 I(m^2)\,,
\ee
which, after the inclusion of the tree-level term, and use of \1eq{IM}, becomes 
\begin{equation}
V_{\Gamma^{(0)}}(0) = 1 + V_{\Gamma^{(0)}}^{(1)}(0)
= 1-\frac{\alpha_s}{2\pi} \left[\left(\frac2{\epsilon} -\gamma\right) - \ln (m^2/\mu^2)\right],
\label{RGz0}
\end{equation}
and 
\begin{equation}
V_{G}^{(1)}(0)= \frac{3}{16}i g^2 \left[I(m^2)-I(\lambda^2)\right]
= \frac{3\alpha_s}{64\pi}\ln (m^2/\lambda^2).
\label{RG0}
\end{equation}
Evidently, all dependence on $1/\epsilon$ is contained in the  coefficient multiplying $\Gamma^{(0)}$, while 
the coefficient of $G$ is completely free of such terms, exactly as one would 
expect from the renormalizability of the theory. Indeed, given that the term $G$ does not appear in the 
original Lagrangian, a divergence of this type could not be renormalized away. Instead, the divergence proportional 
to $\Gamma^{(0)}$ will be reabsorbed in the standard way, namely through the introduction of the appropriate 
vertex renormalization constant, to be denoted by  $Z_4$.

Specifically, one obtains the renormalized vertex $\Gamma_{\!\s{\rm R}}$ from its 
unrenormalized counterpart $\Gamma_{0}$ through the 
condition  (suppressing all indices)
\be
\Gamma_{\!\s{\rm R}}(p_i)
= Z_4 \Gamma_{0}(p_i).
\ee
Of course, the exact form of the $Z_4$ 
and the resulting  $\Gamma_{\!\s{\rm R}}$ depend on the renormalization scheme chosen.
In particular, in the minimal subtraction (MS) scheme one would simply have 
\be
Z_4^{(\s {\rm MS})} = 1+\frac{\alpha_s}{2\pi} \left(\frac2{\epsilon} -\gamma\right), 
\ee
which, upon multiplication with the $V_{\Gamma^{(0)}}(0)$ of \1eq{RGz0}
will give (keeping up to terms of order $\alpha_s$) the finite result 
\be
V_{\Gamma^{(0)}}^{(\s {\rm MS})}(0) =  1 + \frac{\alpha_s}{2\pi} \ln (m^2/\mu^2). 
\ee
Note that $Z_4^{(\s {\rm MS})}$ coincides with the part proportional to $1/\epsilon$ of the 
corresponding expression given in (3.10) of~\cite{Pascual:1980yu} (in the Landau gauge, and for $N=3$).

If one were instead to renormalize in the minimal subtraction (MOM) scheme, as is customary 
in lattice simulations and SDE studies, one would need to introduce a renormalization point, $\mu_{\s{\rm R}}$, and demand  
that at that point the value of the renormalized vertex reduces to its tree-level value. 
For instance, as in~\cite{Pascual:1980yu}, the completely symmetric  
choice $p_i^2 = \mu^2_{\s{\rm R}}$ and $p_i \cdot p_j =-\mu^2_{\s{\rm R}}/3$ may be employed;
then, the corresponding $Z_4^{(\s {\rm MOM})}$ would read (in general)
\be
Z_4^{(\s {\rm MOM})} = 1 - V_{\Gamma^{(0)}}^{(1)}(\mu^2_{\s{\rm R}})  ,
\label{ZMOM}
\ee
such that (schematically)
\be
V_{\Gamma^{(0)}}^{(\s {\rm MOM})}(p_i^2) = 1 + [V_{\Gamma^{(0)}}^{(1)}(p_i^2) - V_{\Gamma^{(0)}}^{(1)}(\mu^2_{\s{\rm R}})].
\label{VMOM}
\ee

Of course, for the case at hand, since the vertex has been computed only for vanishing momenta, one cannot implement a MOM-type scheme. However, in order to get a sense of the general trend that one might expect from a general calculation, we may assume that the subtraction point lies sufficiently far in the UV. Then, for a representative large Euclidean momentum $P$, the qualitative behaviour of the form factor may be approximated by     
\be 
V_{\Gamma^{(0)}}(P^2) \approx  1-\frac{\alpha_s}{2\pi} \left[\left(\frac2{\epsilon} -\gamma\right) - \ln (P^2/\mu^2)\right],
\ee
so that, at $P^2 = \mu^2_{\s{\rm R}}$ one obtains 
\be
Z_4^{(\chic {\rm MOM})} \approx 1 + \frac{\alpha_s}{2\pi} \left[\left(\frac2{\epsilon} -\gamma\right) - \ln ( \mu^2_{\s{\rm R}}/\mu^2)\right],  
\ee
and therefore, the value of $V_{\Gamma^{(0)}}(0)$ gets renormalized to 
\be
V_{\Gamma^{(0)}}^{(\s {\rm MOM})}(0) \approx 1 + \frac{\alpha_s}{2\pi} \ln (m^2/ \mu^2_{\s{\rm R}}).
\ee
As happens typically, in the finite result the 't Hooft scale $\mu$ has been replaced by the renormalization scale $ \mu_{\s{\rm R}}$.

It is obvious at this point, that the above approximations require that  $ \mu^2_{\s{\rm R}} > m^2$, and, consequently, since the logarithm becomes negative, $V_{\Gamma^{(0)}}^{(\s {\rm MOM})}(0) < 1$. To obtain a quantitative notion of the effect, we will use lattice-inspired values for  $m^2$ and $\mu^2_{\s{\rm R}}$; specifically, if we identify the saturation point of the gluon propagator on the lattice with  $1/m^2$, we know that, for $\mu_{\s{\rm R}} = 4.3$ GeV we have that  $m=375$ MeV. Then, using that, for this particular $\mu_{\s{\rm R}}$, $\alpha_s \approx 0.22$, we finally find 
\be
V_{\Gamma^{(0)}}^{(\s {\rm MOM})}(0) \approx  0.83.
\label{perpred}
\ee
Quite interestingly, this apparent tendency of the quantum corrections to reduce the tree-level value persists in the full one-loop dressed calculation; in fact, the value quoted in \1eq{perpred} is fairly close to the one found in the next section.
 
Turning to the $V_{G}^{(1)}(0)$ in \1eq{RG0}, we notice that, when the artificial IR cutoff $\lambda$ is taken to zero, while the physical gluon mass is kept at a nonvanishing value, the logarithm diverges to $+\infty$. Again, this coincides with the behavior found in the more complete calculation of the next section. Of course, the slope of the logarithm found in~\1eq{RG0} is numerically rather suppressed when compared to the result found in the next section; however, this is to be expected, given that the function $F(k^2)$, which in~\1eq{ghostboxes} is raised to the fourth power, is considerably different from 1 in the IR and intermediate momenta.   

\section{\label{sec:nonzeroext}The special momentum configuration $(p,p,p,-3p)$}

Even within the one-loop dressed approximation we are employing, the calculation of the four-gluon vertex for a generic external momenta configuration (such as the one depicted in~\fig{fig:4g-tree-level}) is still a complex task. In addition, it is not the most expeditious way to obtain information about the IR dynamics of this vertex that could be easily contrasted with lattice simulations. 

Thus, we will study a relatively simple kinematic configuration, which
is obtained choosing  a single momentum scale $p$  and identifying the
momentum flow  (see~\fig{fig:4g-tree-level}) with $p_1=p_2=p_3=p$ (and
hence  $p_4=-3p$).   This kinematic  configuration  
gives rise  to  loop integrals  that are  fully
symmetric under the  crossing of external legs; therefore, 
the crossed diagrams may be obtained from the original ones through simple 
permutations of the color and Lorentz indices.

As before, we  will only consider  terms that are quadratic  in the
metric  $gg$. This choice,  in addition  to simplifying  the algebraic
structures  considerably, corresponds  precisely to  the contributions
that  would  survive on  the  lattice, if  one  were  to consider  the
standard quantities  employed in the simulations of  vertices~\cite{Cucchieri:2006tf,  Cucchieri:2008qm}  
(we will return to this point in Sect.~\ref{sec:lattice}).

\subsection{\label{sec:analytic} Analytical results}

Consider the contribution of the ghost boxes.   
The aforementioned crossing property implies that the six different diagrams are proportional to the same integral. 
As a result, one obtains, similarly to what happens in the $p=0$ case, 
\begin{equation}
\sum_{i=1}^6\left.(a_i)^{abcd}_{\mu\nu\rho\sigma}\right\vert_{gg}=g^2\G A(p^2),
\end{equation}
where now
\begin{equation}
A(p^2)=-\frac92\frac1{d^2-1}\kint k^2\left[1-\frac{(k\cd p)^2}{k^2p^2}\right]^2
\frac{F(k)F(k+p)F(k+2p)F(k+3p)}{(k+p)^2(k+2p)^2(k+3p)^2}.
\label{Ap}
\end{equation}
It can be easily checked that as $p\to0$, $A(p^2)$ above reduces to the $A(0)$ of~\1eq{ghostboxes}; therefore we expect that the $\G$ form factor will develop a (logarithmic) divergence in the deep IR.

Next, we consider the gluon boxes. The uncrossed diagram shown in~\fig{fig:4g-1loop-dressed}, yields the general expression
\begin{align}
\left.(b_1)^{abcd}_{\mu\nu\rho\sigma}\right\vert_{gg}&=16 g^2\Tra(T^aT^bT^cT^d)I_{\mu\nu\rho\sigma}(p^2),\nonumber \\
I_{\mu\nu\rho\sigma}(p^2)&=I_1(p^2)g_{\mu\nu}g_{\rho\sigma}+I_2(p^2)g_{\mu\rho}g_{\nu\sigma}+I_3(p^2)g_{\mu\sigma}g_{\nu\rho}+I_4(p^2)\R,
\end{align}
where the integrals $I_i(p^2)$ are not needed for the moment. Crossed diagrams are then obtained from the above expression through the 
replacements $(\mu\nu\rho\sigma)\to(\mu\rho\nu\sigma)$, $(abcd)\to(acbd)$ and $(\mu\nu\rho\sigma)\to(\nu\mu\rho\sigma)$, $(abcd)\to(bacd)$. 
In addition, it turns out that the integrals $I_1$ and $I_3$ are equal\footnote{Notice that without this equality the gluon box contributions would 
lie outside the subset of all possible color and Lorentz tensor structures spanned by $\Gz$ and $\G$. 
Moreover, observe that the realization of this 
equality requires shifts of the integration variable of the type $k \to k+p$; of course, 
since only terms quadratic in the metric are kept, one consistently drops in the numerators terms produced by these shifts that are 
proportional to  $p$ and carry  Lorentz indices of the external legs.} 
upon the momentum shifting $k+3p\to-k$, so that $I_{\mu\nu\rho\sigma}$ can be cast in the form
\begin{equation}
I_{\mu\nu\rho\sigma}(p^2)=[I_2(p^2)-I_1(p^2)]g_{\mu\rho}g_{\nu\sigma}+[I_1(p^2)+I_4(p^2)]\R.
\end{equation}
Thus, adding the three diagrams, one obtains
\begin{equation}
\sum_{i=1}^3\left.(b_i)^{abcd}_{\mu\nu\rho\sigma}\right\vert_{gg}=g^2\Gz B_1(p^2)+g^2\G B_2(p^2),
\end{equation}   
where
\begin{equation}
B_i(p^2)=\kint f_i(k,p)\frac{\Delta(k)\Delta(k+p)\Delta(k+2p)\Delta(k+3p)}{k^2(k+p)^2(k+2p)^2(k+3p)^2},
\end{equation}
and the functions $f_i(k,p)$ are reported in~\1eq{app:glbox}.

\begin{figure}[!t]
\includegraphics[scale=0.775]{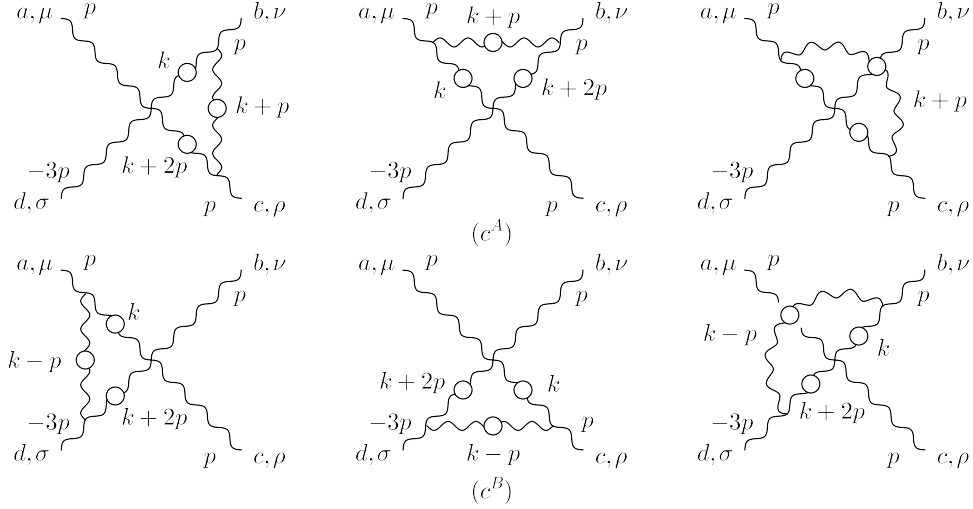}\caption{\label{fig:triangles-1loop-dressed}The 6 one-loop dressed triangle diagrams subdivided in two classes, containing three diagrams each, proportional to independent momentum integrals.}
\end{figure}

We next consider the triangle diagrams. In this case the six graphs can be divided in 
two separate classes (see~\fig{fig:triangles-1loop-dressed}), proportional 
to two independent momentum integrals, namely 
$\Delta(k)\Delta(k+p)\Delta(k+2p)$ [class $(c^\s{A})$] and $\Delta(k)\Delta(k-p)\Delta(k+2p)$ [class $(c^\s{B})$].
Let us then start from the first diagram of the $(c^\s{A})$ class (see again~\fig{fig:triangles-1loop-dressed}); one obtains the general result
\begin{align}
\left.(c^{\s{A}}_1)^{abcd}_{\mu\nu\rho\sigma}\right\vert_{gg}&=6g^2f^{adx}f^{bcx}J_{\mu\nu\rho\sigma}(p^2)-4g^2\Tra(T^aT^bT^cT^d)K_{\mu\nu\rho\sigma}(p^2)\nonumber\\
&-4g^2\Tra(T^aT^dT^bT^c)K_{\mu\rho\nu\sigma}(p^2),
\end{align}
where 
\begin{align}
J_{\mu\nu\rho\sigma}(p^2)&=J_1(p^2)(g_{\mu\nu}g_{\rho\sigma}-g_{\mu\rho}g_{\nu\sigma}),
%this is K1 in the codes
\nonumber \\
K_{\mu\nu\rho\sigma}(p^2)&=K_1(p^2)g_{\mu\sigma}g_{\nu\rho}+K_2(p^2)g_{\mu\rho}g_{\nu\sigma}+K_3(p^2)g_{\mu\nu}g_{\rho\sigma}+K_4(p^2)\R,
%in the codes these are K2, K3, K4 and K5
\end{align}
where again $J_i(p^2)$ and $K_i(p^2)$ are integrals whose explicit expression is not needed at this point.

Within this class, the remaining diagrams are then obtained through the replacements $(\mu\nu\rho\sigma)\to(\sigma\mu\nu\rho)$, $(abcd)\to(dabc)$ and $(\mu\nu\rho\sigma)\to(\sigma\mu\rho\nu)$, $(abcd)\to(dabc)$. Thus, summing up all the $(c^{\s{A}})$ graphs, one obtains, similarly to the zero external momentum case~\1eq{zero-3g}, the result
\begin{equation}
\sum_{i=1}^3\left.(c^{\s{A}}_i)^{abcd}_{\mu\nu\rho\sigma}\right\vert_{gg}=g^2\Gz C_1(p^2)+g^2\G C_2(p^2),
\end{equation}
where
\begin{equation}
C_i(p^2)=\kint g_i(k,p)\frac{\Delta(k)\Delta(k+p)\Delta(k+2p)}{k^2(k+p)^2(k+2p)^2},
\end{equation}
with the $g_i(k,p)$ functions given in~\1eq{app:gltri-ca}. 

Similarly, for the $(c^{\s{B}})$ class we obtain
\begin{equation}
\sum_{i=1}^3\left.(c^{\s{B}}_i)^{abcd}_{\mu\nu\rho\sigma}\right\vert_{gg}=g^2\Gz C'_1(p^2)+g^2\G C'_2(p^2),
\end{equation}
where now
\begin{equation}
C'_i(p^2)=\kint g'_i(k,p)\frac{\Delta(k)\Delta(k-p)\Delta(k+2p)}{k^2(k-p)^2(k+2p)^2},
\end{equation}
and the $g'_i(k,p)$ functions given in~\1eq{app:gltri-cb}. 

We are finally left with the fishnet diagrams. The uncrossed diagram of~\fig{fig:4g-1loop-dressed} yields
\begin{align}
\left.(d_1)^{abcd}_{\mu\nu\rho\sigma}\right\vert_{gg}&=6g^2 f^{adx}f^{bcx}H_{\mu\nu\rho\sigma}(p^2)+g^2\Tra(T^aT^bT^cT^d)L_{\mu\nu\rho\sigma}(p^2)\nonumber\\
&+g^2 \Tra(T^aT^dT^bT^c)L_{\mu\rho\nu\sigma}(p^2),
\end{align}
where 
\begin{align}
H_{\mu\nu\rho\sigma}(p^2)&=H_1(p^2)(g_{\mu\nu}g_{\rho\sigma}-g_{\mu\rho}g_{\nu\sigma}),
%this is L1 in the codes
\nonumber \\
L_{\mu\nu\rho\sigma}(p^2)&=L_1(p^2)g_{\mu\sigma}g_{\nu\rho}+L_2(p^2)g_{\mu\nu}g_{\rho\sigma}+L_3(p^2)\R.
%in the codes these are L2, L4 and L5
\end{align}  

The crossed diagrams are next obtained from the above result through the replacement rules $(\mu\nu\rho\sigma)\to(\mu\nu\sigma\rho)$, $(abcd)\to(adcb)$ and $(\mu\nu\rho\sigma)\to(\mu\sigma\rho\nu)$, $(abcd)\to(dabc)$. Then, summing up all diagrams, one obtains
\begin{equation}
\sum_{i=1}^3\left.(d_i)^{abcd}_{\mu\nu\rho\sigma}\right\vert_{gg}=g^2 \Gz D_1(p^2)+g^2 \G D_2(p^2),
\end{equation}
where now
\begin{equation}
D_i(p^2)=\kint h_i(k,p)\frac{\Delta(k)\Delta(k+2p)}{k^2(k+2p)^2},
\end{equation}
where the $h_i(k,p)$ functions are given in~\1eq{app:glfish}.  

At this point, using the above results, and taking into account the definition~\noeq{4g-general}, one has that
the four-gluon vertex can be cast in the form
\be
\left.\Gamma_{\mu\nu\rho\sigma}^{abcd}(p,p,p,-3p)\right\vert_{gg} =  V_{\Gamma^{(0)}}(p^2) \Gz + V_{G}(p^2) \G,
\ee
with  
\begin{align}
V_{\Gamma^{(0)}}(p^2)&=1+4\pi i\alpha_s[B_1(p^2)+C_1(p^2)+C_1'(p^2)+D_1(p^2)],\nonumber \\
V_{G}(p^2)&=4\pi i\alpha_s\left[A(p^2)+B_2(p^2)+C_2(p^2)+C_2'(p^2)+D_2(p^2)\right],
\label{RGz-RG}
\end{align}
where the ``1'' in  $V_{\Gamma^{(0)}}(p^2)$ represents the tree-level contribution.

\subsection{\label{sec:numeric}Numerical results}

\begin{figure}[!t]
\centerline{\includegraphics[scale=1]{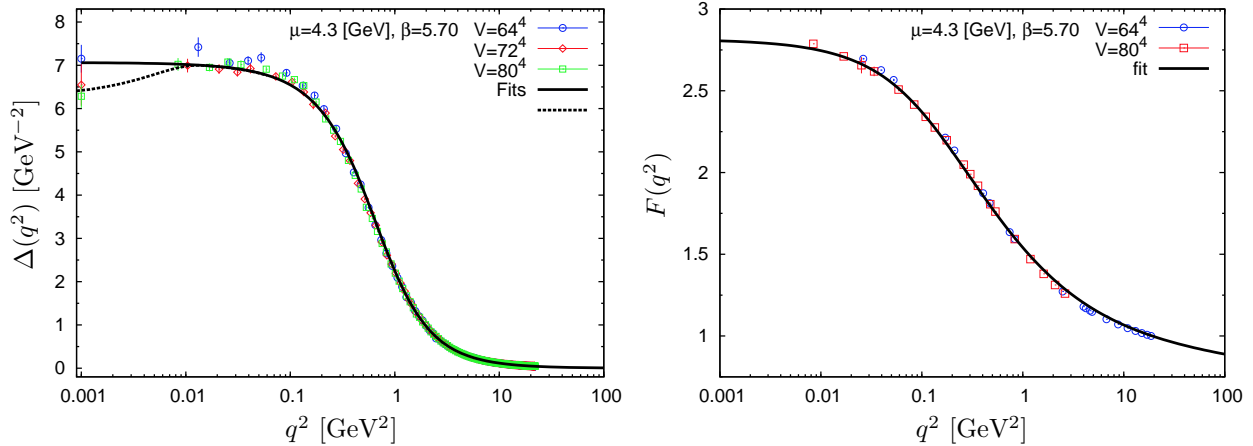}}
\caption{\label{fig:lattice-input} (color online). The SU(3) gluon propagator (left) and ghost dressing function (right) evaluated on the lattice~\cite{Bogolubsky:2009dc} and the corresponding physically motivated fits we use~\cite{Aguilar:2010gm}. In the case of the gluon propagator the dashed curve shows a fit featuring an inflection point the origin of which is linked to the presence of ghost loops~\cite{Aguilar:2013vaa}. All functions are renormalized at $\mu=4.3$~GeV.}  
\end{figure}

In order to study numerically the various one-loop dressed contributions to the four-gluon vertex, 
let us first pass to Euclidean space by defining $k^0\to i\uE{k}_4$ and
$k^j\to-\uE{k}_j$, from which the replacement rules 
${\rm d}^4k\to i{\rm d}^4\dE{k}$, $k\cdot q\to -\dE{k}\cdot\dE{q}$ and $k^2\to-\dE{k}^2$ follow. Next, we introduce spherical coordinates, setting 
\begin{align}
& x=p^2;\qquad y=k^2;\qquad z_n=(k +n p)^2=n^2x+y+2n\sqrt{xy}\cos\theta;\nonumber\\
& \int_{k_\s{\mathrm{E}}}=\frac1{(2\pi)^3}\int_0^\pi\!\mathrm{d}\theta\,\sin^2\theta\int_0^\infty\!\mathrm{d}y\,y.
\end{align}
At this point, all the integrals 
derived in our analytical calculation may  
be evaluated by standard integration techniques, provided that 
we supply as input the gluon propagator $\Delta$ and the ghost dressing function $F$. 

\begin{figure}
\includegraphics[scale=0.85]{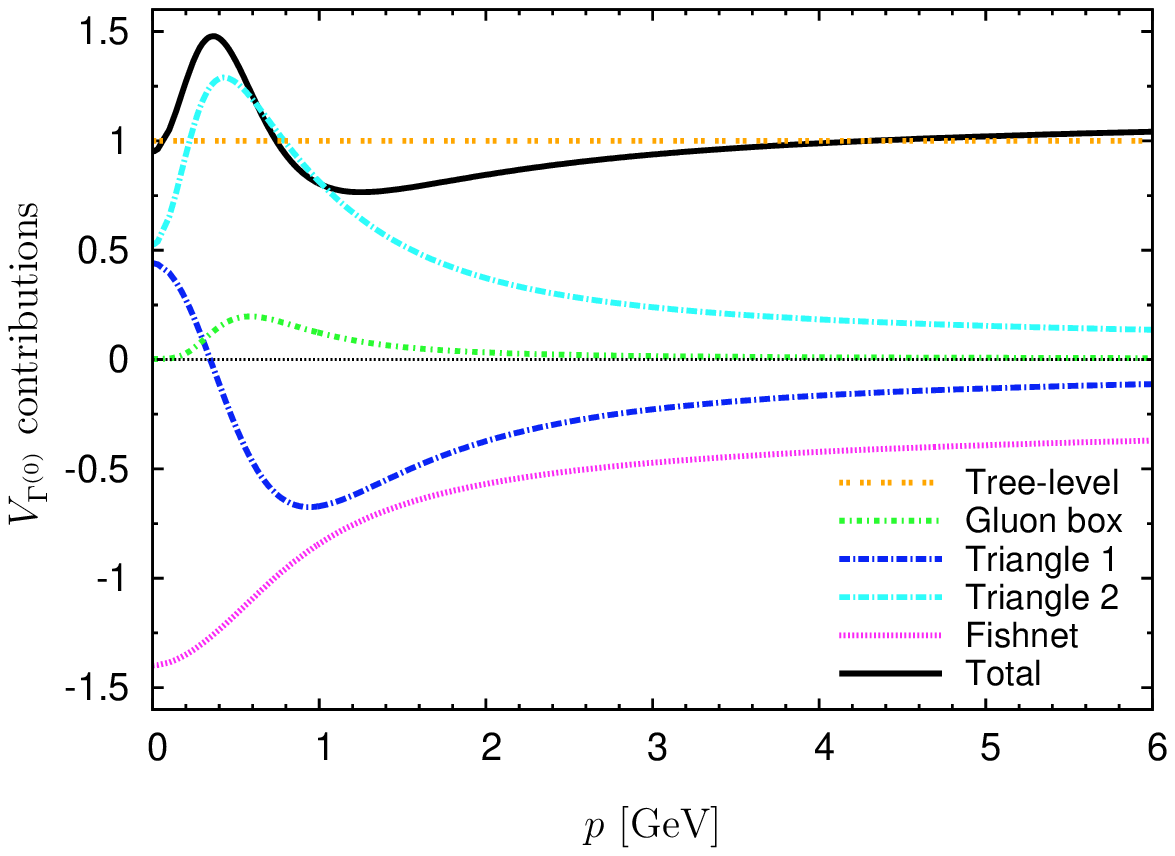}
\caption{\label{fig:Gamma0} (color online). Individual  one-loop dressed contributions to the tensor structure $\Gz$. The black line coincides with the coefficient $V_{\Gamma^{(0)}}$ of~\1eq{RGz-RG}.}  
\end{figure}

To this end, we use physically motivated fits to the lattice data of~\cite{Bogolubsky:2007ud}, 
whose explicit functional form can be found in~\cite{Aguilar:2010gm}. 
The agreement of these fits with the corresponding lattice data at the renormalization scale $\mu=4.3$ GeV is shown in~\fig{fig:lattice-input}. 
For the case of the gluon propagator we also show a fit displaying 
the inflection point that must appear due to the presence of divergent ghost loops~\cite{Aguilar:2013vaa}; 
the results obtained are practically independent from the implementation of this feature in the gluon propagator.

\begin{figure}
\includegraphics[scale=0.85]{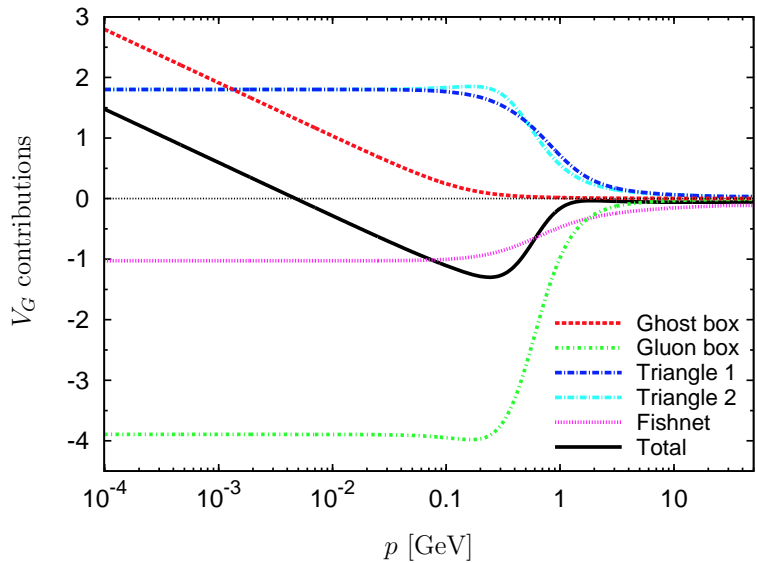}
\caption{\label{fig:G} (color online). Individual one-loop dressed contributions to the tensor structure $\G$. The black line coincides with the coefficient $V_G$ of~\1eq{RGz-RG}.}  
\end{figure}

In~\fig{fig:Gamma0} and~\fig{fig:G} we plot, respectively, the contributions of the various diagrams to $V_{\Gamma^{(0)}}$ and $V_G$, 
together with their total sum (in the former case, all terms have been subtractively renormalized within the MOM scheme, at $\mu=4.3$ GeV, 
in accordance with \1eq{VMOM}). 

As already mentioned, ghost boxes will not contribute to $V_{\Gamma^{(0)}}$, which is entirely made up of gluonic contributions, 
all of them saturating in the IR~(see \fig{fig:Gamma0}, again). The contribution of the gluon boxes is negligible; indeed, as $p\to0$ it vanishes, 
as we know it should from the zero external momenta case (see Table~\ref{tab:1lres}). The triangle terms feature a bump 
of opposite sign, while the fishnet is negative. Adding everything up, one obtains the shape shown by the black line of~\fig{fig:Gamma0}.
 Notice that at zero momentum we obtain the value $V^{(\s{\rm MOM})}_{\Gamma^{(0)}}=0.95$, 
which compares rather well with the perturbative estimate of~\1eq{perpred}.

In the case of $V_G$ the situation is completely different (\fig{fig:G}). Gluon contributions are again saturating in the IR; 
however, in this case, the ghost boxes take over below few hundreds MeV$^2$, driving $V_G$ to an IR logarithmic divergence. 
In fact, the IR behavior is perfectly described by the function $a\log x+b$ with $a=-0.187$ and $b=- 1.989$. As far as the remaining diagrams are concerned, gluon boxes are negative in this case; in addition, they are almost perfectly cancelled by the two triangle contributions, 
which (contrary to the previous case) have now the same sign. 
When the negative contribution from the fishnet diagrams is finally added, one obtains the shape shown by the black line of~\fig{fig:G}. 

It is important to notice that $V_G$ displays a zero crossing, a feature that is also present in the $R$ ratio defined in the case of the three-gluon vertex~\cite{Aguilar:2013vaa,Blum:2014gna,Eichmann:2014xya}. The location of the crossing appears to be very deep in the IR (around a few MeV); recall, however, 
that the $V_G$ displayed in~\fig{fig:G} has been evaluated without dressing the various vertices. 
In order to obtain an estimate of the possible impact that vertex corrections might have on the behavior of $V_G$, 
let us consider what happens when the ghost vertices, appearing in the ghost box diagrams $(a)$, are dressed. 

Writing for the ghost vertex (all momenta entering)
\begin{equation}
i\Gamma_{c^n A^a_\alpha \bar c^m}(k-p,p,-k)=gf^{amn}\Gamma_\alpha(k-p,p,-k),
\end{equation}
the most general tensorial structure decomposition of $\Gamma_\alpha$ is given by
\begin{align}
\Gamma_\alpha(k-p,p,-k)&={\cal A}(k-p,p,-k)k_\alpha+{\cal B}(k-p,p,-k)p_\alpha.
\label{gh-vert-deco}
\end{align}
Evidently, at tree-level, one has ${\cal A}^{(0)}=1$ and ${\cal B}^{(0)}=0$. 

\begin{figure}[!t]
\centerline{\includegraphics[scale=1]{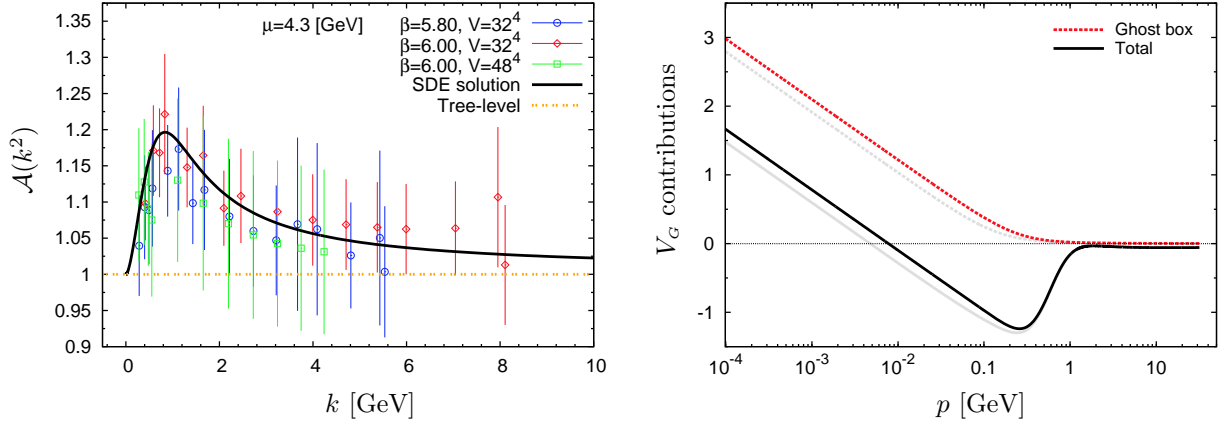}}
\caption{\label{fig:Ghost-vertex-corr} (color online). {\it Left panel.} The ghost vertex form factor ${\cal A}$ in the soft gluon limit calculated in the continuum and on the lattice. {\it Right panel.} The ghost contribution to $V_G$, together with the total $V_G$, 
when ghost vertex corrections are included. The gray lines represent the same quantities evaluated with tree-level vertices: 
the shift of the zero crossing towards the right is evident.}  
\end{figure}

To be sure, the form factors ${\cal A}$, ${\cal B}$ are not known for arbitrary momenta. 
However, in the soft gluon limit ($p\to0$), which is the most relevant to our purposes, 
\1eq{gh-vert-deco} reduces to
\begin{align}
\Gamma_\alpha(k,0,-k)&={\cal A}(k^2)k_\alpha;& {\cal A}(k^2)={\cal A}(k,0,-k), 
\label{gh-vert-soft}
\end{align}
and the form factor ${\cal A}(k^2)$
has been studied both in the continuum~\cite{Boucaud:2011eh,Dudal:2012zx,Aguilar:2013xqa}
and on the lattice, both for SU(2)~\cite{Cucchieri:2004sq} and  SU(3)~\cite{Sternbeck:2006rd}; 
the results obtained in~\cite{Aguilar:2013xqa} and~\cite{Sternbeck:2006rd}
are shown in the left panel of~\fig{fig:Ghost-vertex-corr}. 
As can be seen, in this limit ${\cal A}$ develops a sizeable peak around 800 MeV, approaching 
its tree-level value for both IR as well as UV momenta.

The idea is then to replace all ghost-gluon vertices appearing in a generic ghost box diagram by~\1eq{gh-vert-soft}; 
this amounts to effectively multiplying the integrand in~\1eq{Ap} by ${\cal A}^4(k^2)$. 
Obviously this operation constitutes an approximation, which is, nevertheless, 
reliable in the IR momentum region. 
The resulting modification of the ghost box contribution to $V_G$, as well as the total $V_G$, are then shown on the right panel of~\fig{fig:Ghost-vertex-corr}. The inclusion of vertex corrections causes a change in the logarithmic IR running (with now $a=-0.191$ and $b=- 1.858$) of the ghost contribution, to which corresponds a shift of the zero crossing point of $V_G$ towards higher momentum (one gets twice the value obtained with tree-level ghost vertices). 
It remains to be seen if the addition of vertex corrections to all other diagrams will produce a further shift of the point of the zero crossing 
towards the moderate momentum region, of say $p\sim 0.1\div1$ GeV, where the onset 
of nonperturbative effects appears to take place in all previously studied SU(3) Green's functions.

\subsection{\label{sec:lattice}Lattice quantities}

We conclude this section by commenting on certain issues that appear when the quantities  defined on the lattice 
for studying vertices~\cite{Cucchieri:2006tf,  Cucchieri:2008qm} (to date, only three-point functions) 
are extended to the case of the four-gluon vertex. 

\begin{figure}[!t]
%\mbox{}\hspace{0cm}
\includegraphics[scale=0.725]{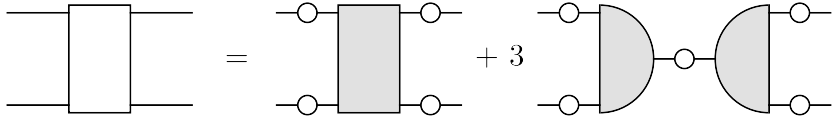} 
\caption{\label{fig:connected}
Schematic decomposition of a four-point connected Green's function into its 1-PI and 1-PR contributions. The factor of 3 takes into account crossed diagrams. White (respectively, gray) boxes/blobs represent connected (respectively 1-PI) functions.}
\end{figure}

Lattice simulations of the four gluon vertex would be challenging for the following two reasons. On 
the one hand, simulations of multi-gluons correlation functions are noisy, therefore requiring the sampling of a very large number of gauge configurations. 
On the other hand, lattice calculations are bound to probe the connected Green's functions ${\cal C}$ rather than the 1-PI
functions $\Gamma$. In the four gluon case addressed in this paper, the two functions are related by (see also~\fig{fig:connected})
\begin{align}
{\cal C}^{abcd}_{\mu\nu\rho\sigma}(p_1,p_2,p_3,p_4)&=\Delta^{\mu\alpha}(p_1)\Delta^{\nu\beta}(p_2)\Delta^{\rho\gamma}(p_3)\Delta^{\sigma\delta}(p_4)\Gamma^{abcd}_{\alpha\beta\gamma\delta}(p_1,p_2,p_3,p_4)-i\Delta^{\mu\alpha}(p_1)\Delta^{\sigma\delta}(p_4)\times\nonumber \\
&\times\Gamma^{mad}_{\varepsilon\alpha\delta}(p_1+p_4,p_1,p_4)\Delta^{\varepsilon\varepsilon'}(p_1+p_4)\Gamma^{mbc}_{\varepsilon'\beta\gamma}(p_2+p_3,p_2,p_3)
\Delta^{\nu\beta}(p_2)\Delta^{\rho\gamma}(p_3)\nonumber\\
&+{\rm crossing} \ {\rm terms},
\label{conn}
\end{align}
and we see that 1-PR diagrams constructed from lower order functions will spoil in general the possibility of isolating the genuine 1-PI contribution to ${\cal C}$. 

In order to address the second problem, 
observe that, in the Landau gauge, the only rank-2 Minkowski tensor allowed is the transverse projector, while, in general, the allowed rank-3 tensors for the three-gluon vertex are either 
linear in both the metric and momenta ($gp$), or cubic in momenta ($pqr$). This means, in turn, that 
in the momentum configuration $(p_1,p_2,p_3,p_4)=(p,p,p,-3p)$, each propagator appearing in the decomposition~\1eq{conn} will have an accompanying projector $P(p)$;  as a result, all 1-PR contributions will vanish, and one is left with 
\begin{equation}
{\cal C}^{abcd}_{\mu\nu\rho\sigma}(p,p,p,-3p)=\Delta^3(p^2)\Delta(9p^2)P^{\mu\alpha}(p)P^{\nu\beta}(p)P^{\rho\gamma}(p)P^{\sigma\delta}(p)\left.\Gamma^{abcd}_{\alpha\beta\gamma\delta}(p,p,p,-3p)\right\vert_{gg}.
\end{equation}
Therefore, we arrive at the important conclusion that the momentum configuration $(p,p,p,-3p)$ allows the study of the (projected) 1-PI component of the four-gluon connected Green's function in isolation. 

At this point, the scalar factors $\Delta$ can be factored out by defining the lattice $R$ ratio in the standard way~\cite{Cucchieri:2006tf,Cucchieri:2008qm}, namely projecting ${\cal C}$ on a suitable tensor $\T$, and normalizing the resulting expression. Specifically, one writes
\begin{equation}
R_T(p^2)=\frac{{\T\cal C}_{abcd}^{\mu\nu\rho\sigma}}{\T P^{\mu\alpha}P^{\nu\beta}P^{\rho\gamma}P^{\sigma\delta}T^{abcd}_{\alpha\beta\gamma\delta}}
=\frac{\T P^{\mu\alpha}P^{\nu\beta}P^{\rho\gamma}P^{\sigma\delta}\left.\Gamma^{abcd}_{\alpha\beta\gamma\delta}\right\vert_{gg}}{\T P^{\mu\alpha}P^{\nu\beta}P^{\rho\gamma}P^{\sigma\delta}T^{abcd}_{\alpha\beta\gamma\delta}}.
%\Gamma^{abcd(0)}_{\alpha\beta\gamma\delta}}.
\label{RT}
\end{equation}
One then usually chooses $T$ to coincide with the tree-level vertex $\Gamma^{(0)}$, 
so that any deviation of $R$ from 1 signals the onset of quantum (nonperturbative) effects. 

For the case of the four gluon vertex, however, additional care is needed, depending on the particular 
property that one attempts to expose\footnote{In particular, if one wants to capture the aforementioned logarithmic 
IR divergence attributed to the massless ghost loops, the orthogonality relations~\noeq{ortho} implies that $T\neq \Gamma^{(0)}$.}. 
Indeed, the general analysis based on Bose symmetry of Appendix~\ref{Bose}, reveals that in the momentum configuration under 
scrutiny there are at most three possible tensor structures (proportional to $gg$) contributing to 
the {\it full} four-gluon vertex, \ie one has\footnote{Obviously, within the one-loop approximation 
that we have employed in our calculations, one has $V_{X'}=0$, so that~\1eq{ortho} implies the identifications $V_{\Gamma^{(0)}}=R_{\Gamma^{(0)}}$ and $V_G=R_G$ 
($R_{X'}$ is redundant in this case).}
\begin{equation}
\left.\Gamma^{abcd}_{\mu\nu\rho\sigma}(p,p,p,-3p)\right\vert_{gg}=V_{\Gamma^{(0)}}(p^2)\Gz+V_G(p^2)\G+V_{X'}(p^2)X_{\mu\nu\rho\sigma}^{'abcd}.
\label{fulldeco}
\end{equation}
Thus, the complete structure of the four gluon vertex can be obtained by defining the three different ratios corresponding to setting $T=\Gamma^{(0)}$, $G$ and $X'$ in the definition~\noeq{RT}.

In addition, as explained in Appendix~\ref{Bose}, Bose symmetry alone does not unambiguously fix the 
tensor $X'$, whose exact form depends on the choice of the ``basis'' that spans  this particular space.  
The various possible choices are conveniently parametrized by means of a parameter $s$ [see~\1eq{Xprime}]. 

Ideally one would like to fix $s$ in a way such that 
the resulting $X'$ be orthogonal to both $\Gamma^{(0)}$ and $G$, that is by requiring
\begin{equation}
\Gz P^{\mu\alpha}(p)P^{\nu\beta}(p)P^{\rho\gamma}(p)P^{\sigma\delta}(p)X^{'abcd}_{\alpha\beta\rho\gamma}=\G P^{\mu\alpha}(p)P^{\nu\beta}(p)P^{\rho\gamma}(p)P^{\sigma\delta}(p)X^{'abcd}_{\alpha\beta\rho\gamma}=0.
\end{equation}
This is however not possible, as $X'$ can be rendered orthogonal to either $\Gamma^{(0)}$ (for $s=0$) or $G$ (for $s=1/3$), but not both. In these cases one has
\begin{align}
&s=0:&R_{\Gamma^{(0)}}&=V_{\Gamma^{(0)}};& R_G&=V_G+\frac19V_{X'};& R_{X'}&=V_{X'}+\frac3{13}V_G,\nonumber\\
&s=\frac13:&R_{\Gamma^{(0)}}&=V_{\Gamma^{(0)}}+\frac1{81}V_{X'};& R_G&=V_G;& R_{X'}&=V_{X'}+\frac9{164}V_{\Gamma^{(0)}}.
\label{project}
\end{align}

There are at least two reasons to prefer the second choice over the first. 
To begin with, recall that, according to our general analysis, the origin of the divergence in the vertex form factors 
is clearly associated with the masslessness of the full ghost propagator; consequently, $V_G$ will continue to be divergent 
even in the full nonperturbative setting provided by a lattice calculation. At the same time, we expect $V_{\Gamma^{(0)}}$ 
to be finite, as the diagrams contributing to it are ``protected'' by the effective gluon mass;
this means, in turn, that in the $s=0$ basis both $R_G$ and $R_{X'}$ will diverge, even in the case of a finite $V_{X'}$. 
In the $s=1/3$ basis, a lattice calculation would instead find that the only IR divergent ratio would be $R_G$, 
immediately signalling a finite $V_{\Gamma^{(0)}}$ and $V_{X'}$ form factors.
In addition, observe that~\1eq{project} implies that the vectors $\Gamma^{(0)}$ and $X'$ are very close to be orthogonal for the $s=1/3$ case; 
for example, when projecting the full vertex along $\Gamma^{(0)}$, the $X'$ component is two orders of magnitude smaller than the $\Gamma^{(0)}$ one, and vice-versa.

A lattice measurement of the ratios $R_T$ in the $s=1/3$ basis will finally yield the complete vertex form factors $V_T$, through the formulas
\begin{align}
V_{\Gamma^{(0)}}&=\frac{164}{13275}\left(81R_{\Gamma^{(0)}}-R_{X'}\right);& V_G&=R_G;& V_{X'}&=\frac9{1475}\left(164R_{X'}-9R_{\Gamma^{(0)}}\right).
\end{align}

\section{\label{concl}Conclusions}

In this paper we have explored certain nonperturbative features of the
SU(3) four-gluon vertex in the Landau gauge. In particular, of the set
of  all possible quantum  corrections, we  have considered  the subset
corresponding to the one-loop  dressed diagrams, in which vertices are
kept at tree-level while propagators  are fully dressed. If only terms
quadratic  in the metric  are kept,  and the  kinematical configuration
$(p,p,p,-3p)$  is chosen,  we have  found that,  within this  class of
diagrams, only two orthogonal  Lorentz and color tensor structures can
emerge: the tree-level vertex $\Gamma^{(0)}$, and the tensor $G$.

It  turns out  that  ghost boxes  can  contribute only  to the  latter
structure, while  all the  remaining diagrams (gluon  boxes, triangle,
and  fishnet diagrams) contributes  to both.  Then, as  massless ghost
loops  invariably lead  to the  presence of  an IR  divergence  in the
corresponding  diagram~\cite{Aguilar:2013vaa},  one  expects the  form
factor $V_G$  (respectively, $V_{\Gamma^{(0)}}$) to be  IR divergent (respectively,
finite).  A numerical study  performed, using  as input  the available
lattice  data for the  gluon propagator  and ghost  dressing function,
confirms these expectations. In addition, one finds that $V_G$ shows a
zero  crossing  before the  form  factor  diverges logarithmically  to
$+\infty$.

It would  certainly be interesting  to scrutinize this  issue further,
and reach  a definite  conclusion on the  way that this  particular IR
divergence manifests itself at the level of the four-gluon vertex. One
possible direction  has already been  pursued here to some  limited extent,
namely the  dressing of the  ghost vertices appearing in  the one-loop
dressed diagrams.  One may attempt to complete this task, by also dressing
the three-gluon vertices;  to be  sure,  
the tensorial  structure  of the  three-gluon
vertices is bound to lead to a proliferation of terms, which, however,
may become manageable in the limit of interest, namely as all external
momenta tend to zero. Even if  this approach would not exhaust all the possible
vertex dressing (the real  problem in this context being  the dressing of the
four-gluon  vertices  appearing   in  the  triangle  diagrams),  the
importance of accomplishing  this step would be twofold:  on the one
hand,  one would  see if  the zero  crossing gets  pushed  towards the
more ``favorable'' momentum region $p\sim  0.1\div1$ GeV,  
as the  ghost vertex  corrections seem to  suggest; on the other hand,  it might be possible
to detect the appearance of 
the  form factor associated to the  $X^{'}$ tensor allowed by
Bose  symmetry,  and  address   its  IR  properties  (or  confirm  its
vanishing).

Of course, the  lattice  could  be  instrumental  in  addressing  all  the
aforementioned   issues.  Indeed,  our   analysis  reveals   that  the
$(p,p,p,-3p)$ configuration would permit the study of the 1-PI part of the
connected four  gluon function alone,  and that the full  structure of
the  vertex can  be then  reconstructed from  the measurements  of the
standard ratios $R_{\Gamma^{(0)}}$, $R_G$ and $R_{X'}$. In that sense, 
the main remaining difficulty to overcome 
is  to  average over  a  large  sample of  gauge
configurations,  in  order to  tame  the  statistical fluctuations.

\acknowledgments 

The research of J.~P. is supported by the Spanish MEYC under grant FPA2011-23596. 

\appendix

\section{\label{Bose}Bose symmetry}

In this Appendix we use the Bose symmetry of the four external gluon legs in order to constrain the number of the possible 
tensors that can appear in the $gg$ part of the four-gluon vertex, specializing to the kinematic configuration $(p_1,p_2,p_3,p_4)=(p,p,p,-3p)$

To begin with, let us classify the possible color structures into three subsets 
\bea
\{A \}&=& \{\delta^{ab} \delta^{cd}, \delta^{ac} \delta^{db},\delta^{ad} \delta^{bc}\},
\nonumber\\
\{B \}&=& \{d^{abr} d^{cdr}, d^{acr} d^{dbr},d^{adr} d^{bcr} \},
\nonumber\\
\{C \} &=& \{d^{abr} f^{cdr}, d^{acr} f^{dbr},d^{adr} f^{bcr} \},
\label{ABC}
\eea
denoting the corresponding elements by
$A_j^{abcd}$, $B_j^{abcd}$, and  $C_j^{abcd}$, with $j=1,2,3$.  
The reason for this particular separation is that the elements of different subsets do 
not mix when one applies the permutations dictated by the Bose symmetry of the four-gluon vertex, 
as we will do below.

In particular, the three basic structures that emerge 
when one considers only terms quadratic in the metric assume the general form 
\bea
(V_1)^{abcd}_{\mu\nu\rho\sigma} &=& g_{\mu\nu}g_{\rho\sigma} [ a_{1j} A_j^{abcd} + b_{1j}B_j^{abcd} + c_{1j}C_j^{abcd}],
\nonumber\\
(V_2)^{abcd}_{\mu\nu\rho\sigma} &=& g_{\mu\rho}g_{\nu\sigma} [ a_{2j} A_j^{abcd} + b_{2j}B_j^{abcd} + c_{2j}C_j^{abcd}],
\nonumber\\
(V_3)^{abcd}_{\mu\nu\rho\sigma} &=& g_{\mu\sigma}g_{\nu\rho} [ a_{3j} A_j^{abcd} + b_{3j}B_j^{abcd} + c_{3j}C_j^{abcd}].
\label{theV}
\eea
and the vertex may be written as 
\be
\Gamma_{\mu\nu\rho\sigma}^{abcd} = \sum_{i=1}^{3}(V_i)^{abcd}_{\mu\nu\rho\sigma}.
\label{GammawithV}
\ee
The coefficients (form factors) $a_{ij}$, $b_{ij}$, and $c_{ij}$ are functions of the 
incoming momenta, namely $a_{ij} = a_{ij}(p_1,p_2,p_3,p_4)$, etc. 
Of course, the Bose symmetry of the four-gluon vertex imposes constraints on their behavior, 
allowing one to express some of them in terms of others (with their arguments permuted), thus reducing the 
number of unknown functions.  The 24 possible permutations correspond to all possible rearrangements of the set $(abcd)$.

\subsection{The case of vanishing momenta}

As a warm up exercise, let us first see what happens to the $V_i$ when all incoming momenta vanish, \mbox{$(p_1,p_2,p_3,p_4)=(0,0,0,0)$}.
In that case, of course, all coefficients  $a_{ij}$, $b_{ij}$ and $c_{ij}$ are simply constants.

Let us first consider the contributions to $V_i$ coming from the first subset, to be denoted by  $V_i^{\chic A}$.
In particular, we have 
\bea
(V_1^{\chic A})^{abcd}_{\mu\nu\rho\sigma} &=& 
g_{\mu\nu}g_{\rho\sigma} [ a_{11} \delta^{ab}\delta^{cd} + a_{12} \delta^{ac}\delta^{bd} + a_{13}\delta^{ad}\delta^{bc} ],
\nonumber\\
(V_2^{\chic A})^{abcd}_{\mu\nu\rho\sigma} &=& g_{\mu\rho}g_{\nu\sigma} [ a_{21} \delta^{ab}\delta^{cd} + a_{22}\delta^{ac}\delta^{bd} + a_{23}\delta^{ad}\delta^{bc}],
\nonumber\\
(V_3^{\chic A})^{abcd}_{\mu\nu\rho\sigma} &=& g_{\mu\sigma}g_{\nu\rho} [ a_{31} \delta^{ab}\delta^{cd} + a_{32}\delta^{ac}\delta^{bd} + a_{33}\delta^{ad}\delta^{bc}].
\label{ViA}
\eea
At this point, the requirement of Bose symmetry under $(a, \mu \leftrightarrow b,\nu)$ 
forces $(V_1^{\chic A})^{abcd}_{\mu\nu\rho\sigma}$
to transform into itself, and as a result we must have  $a_{12} = a_{13}$. Similarly, the requirement of symmetry under $(a, \mu \leftrightarrow c,\rho)$
forces $a_{21} = a_{23}$, whilst symmetry under $(a, \mu \leftrightarrow d,\sigma)$ leads to  $a_{31} = a_{32}$. Thus, the $(V_i^{\chic A})^{abcd}_{\mu\nu\rho\sigma}$
of~\1eq{ViA} reduce to the form 
 \bea
(V_1^{\chic A})^{abcd}_{\mu\nu\rho\sigma} &=& g_{\mu\nu}g_{\rho\sigma} [ a_{11} \delta^{ab}\delta^{cd} + a_{12} (\delta^{ac}\delta^{bd} + \delta^{ad}\delta^{bc}) ],
\nonumber\\
(V_2^{\chic A})^{abcd}_{\mu\nu\rho\sigma} &=& g_{\mu\rho}g_{\nu\sigma} [  a_{22}\delta^{ac}\delta^{bd}+ a_{23} (\delta^{ab}\delta^{cd} + \delta^{ad}\delta^{bc})],
\nonumber\\
(V_3^{\chic A})^{abcd}_{\mu\nu\rho\sigma} &=& g_{\mu\sigma}g_{\nu\rho} [ a_{33}\delta^{ad}\delta^{bc} + a_{31} (\delta^{ab}\delta^{cd} + \delta^{ac}\delta^{bd})].
\eea
Note that if one carries out the second obvious set of permutations, namely $(c,\rho \leftrightarrow d,\sigma)$, 
$(b, \nu \leftrightarrow d,\sigma)$, and $(c, \nu \leftrightarrow b,\rho)$, the  $V_1^{\chic A}$,   $V_2^{\chic A}$, and  $V_3^{\chic A}$, respectively,
are  automatically symmetric. 

Of course, when the permutation is such that one particular $V_i^{\chic A}$ must transform into itself, 
the other two must transform one into the other.
For example, when $(a, \mu \leftrightarrow b,\nu)$, we have that $V_1^{\chic A} \to V_1^{\chic A}$, whereas 
\bea
(V_2^{\chic A})^{abcd}_{\mu\nu\rho\sigma} & \to & g_{\mu\sigma}g_{\nu\rho} [ a_{22}\delta^{ad}\delta^{bc} + a_{23} (\delta^{ab}\delta^{cd} + \delta^{ac}\delta^{bd})],
\nonumber\\
(V_3^{\chic A})^{abcd}_{\mu\nu\rho\sigma} & \to & g_{\mu\rho}g_{\nu\sigma} [  a_{33}\delta^{ac}\delta^{bd}+ a_{31} (\delta^{ab}\delta^{cd} + \delta^{ad}\delta^{bc})].
\eea
Then, Bose symmetry requires that the transformed $V_2^{\chic A}$ must coincide with the original $V_3^{\chic A}$, and vice-versa, and therefore we must have that 
 $a_{22} = a_{33}$, and $a_{23} =  a_{31}$. The repetition of this arguments leads to the conclusion that $a_{11} = a_{22} = a_{33} \equiv {\widehat a}$,
and $a_{12}= a_{23} =  a_{31} \equiv \widetilde a$; thus, finally, after setting $a \equiv {\widehat a}- {\widetilde a}$, we have that  
\bea
(V_1^{\chic A})^{abcd}_{\mu\nu\rho\sigma} &=& g_{\mu\nu}g_{\rho\sigma} [ a \delta^{ab}\delta^{cd} + {\widetilde a} (\delta^{ab}\delta^{cd} + \delta^{ac}\delta^{bd} + \delta^{ad}\delta^{bc}) ],
\nonumber\\
(V_2^{\chic A})^{abcd}_{\mu\nu\rho\sigma} &=& g_{\mu\rho}g_{\nu\sigma} [  a \delta^{ac}\delta^{bd}+  {\widetilde a}(\delta^{ab}\delta^{cd} + \delta^{ac}\delta^{bd}+ \delta^{ad}\delta^{bc})],
\nonumber\\
(V_3^{\chic A})^{abcd}_{\mu\nu\rho\sigma} &=& g_{\mu\sigma}g_{\nu\rho} [ a \delta^{ad}\delta^{bc} + {\widetilde a} (\delta^{ab}\delta^{cd} + \delta^{ac}\delta^{bd}+\delta^{ad}\delta^{bc})].
\label{V1con}
\eea
Note that past this point, use of the remaining possible permutations imposes no further restrictions on the coefficients $a$ and ${\widetilde a}$. 

A completely similar procedure may be applied to the parts of the $(V_i)$ related to the subset $\{B \}$. In particular, one reaches the 
conclusion that 
\bea
(V_1^{\chic B})^{abcd}_{\mu\nu\rho\sigma} &=& g_{\mu\nu}g_{\rho\sigma} [ b d^{abr}d^{cdr} + {\widetilde b} (d^{abr}d^{cdr} + d^{acr}d^{bdr} + d^{adr}d^{bcr}) ],
\nonumber\\
(V_2^{\chic B})^{abcd}_{\mu\nu\rho\sigma} &=& g_{\mu\rho}g_{\nu\sigma} [  b d^{acr}d^{bdr}+  {\widetilde b}(d^{abr}d^{cdr} + d^{acr}d^{bdr}+ d^{adr}d^{bcr})],
\nonumber\\
(V_3^{\chic B})^{abcd}_{\mu\nu\rho\sigma} &=& g_{\mu\sigma}g_{\nu\rho} [ b d^{adr}d^{bcr} + {\widetilde b} (d^{abr}d^{cdr} + d^{acr}d^{bdr}+d^{adr}d^{bcr})].
\label{ViB}
\eea

Turning to the subset $\{C\}$, it is relatively straightforward to establish that it does not contribute to the  $(V_i)$. 
To see this, let us choose any of the $(V_i)$, say $(V_1)$, and consider the general form of its component $(V_1^{\chic C})$, given by 
\be
(V_1^{\chic C})^{abcd}_{\mu\nu\rho\sigma} = g_{\mu\nu}g_{\rho\sigma} [c_{11}d^{abr}f^{cdr} + c_{12}d^{acr}f^{dbr}+c_{13}d^{adr}f^{bcr}].
\ee
Let us now implement the permutation $(c, \rho \leftrightarrow d,\sigma)$, under which 
\be
(V_1^{\chic C})^{abcd}_{\mu\nu\rho\sigma} \to g_{\mu\nu}g_{\rho\sigma} [-c_{11}d^{abr}f^{cdr} - c_{12}d^{adr}f^{bcr} - c_{13}d^{acr}f^{dbr}],
\ee
and since the transformed $(V_1^{\chic C})$ must coincide with the original one, we have that $c_{11}=-c_{11}$ and $c_{12}=-c_{13}$, and so
\bea
(V_1^{\chic C})^{abcd}_{\mu\nu\rho\sigma} &=&  g_{\mu\nu}g_{\rho\sigma} c_{13}[d^{acr}f^{bdr} + d^{adr}f^{bcr}]
\nonumber\\
&=& g_{\mu\nu}g_{\rho\sigma} c_{13} d^{cdr} f^{abr},
\eea
where we have used the second identity of \1eq{ident}.
But this last expression must remain invariant  under the additional permutation  \mbox{$(a, \mu \leftrightarrow b,\nu)$}, which implies that 
$c_{13}=0$. 

A this point one may specialize to the case $N=3$, and use \1eq{N3} into \1eq{ViB}, to write the vertex in the form 
\be
\Gamma_{\mu\nu\rho\sigma}^{abcd}(0,0,0,0) = \left({\widetilde a} + \frac{{\widetilde b} }{3}\right) \G +   L_{\mu\nu\rho\sigma}^{abcd},
\label{Gwith3}
\ee 
with
\begin{align}
L_{\mu\nu\rho\sigma}^{abcd} &=a E_{\mu\nu\rho\sigma}^{abcd}+bE_{\mu\nu\rho\sigma}^{'abcd},&\nonumber \\
E_{\mu\nu\rho\sigma}^{abcd}&=g_{\mu\nu}g_{\rho\sigma}\delta^{ab}\delta^{cd} + g_{\mu\rho}g_{\nu\sigma} \delta^{ac}\delta^{bd} + g_{\mu\sigma}g_{\nu\rho} \delta^{ad}\delta^{bc},\nonumber \\
E_{\mu\nu\rho\sigma}^{'abcd}&=g_{\mu\nu}g_{\rho\sigma}d^{abr}d^{cdr}+g_{\mu\rho}g_{\nu\sigma} d^{acr}d^{bdr}+g_{\mu\sigma}g_{\nu\rho} d^{adr}d^{bcr}.  
\end{align}
The term $L$ may be further manipulated, by noticing that if the condition $a= \frac{2b}{3}$ were satisfied, then we would have that 
$L_{\mu\nu\rho\sigma}^{abcd} = \frac{2b}{3} X_{\mu\nu\rho\sigma}^{abcd} = \frac{b}{3}[\G + \Gz ]$ [see \1eq{gagx}]. 
Therefore, the most general way to rearrange this term is
\begin{align}
L_{\mu\nu\rho\sigma}^{abcd}&=(a-cs)X_{\mu\nu\rho\sigma}^{abcd}+cX_{\mu\nu\rho\sigma}^{'abcd};& X_{\mu\nu\rho\sigma}^{'abcd}&=sE_{\mu\nu\rho\sigma}^{abcd}+\frac32(s-1)E_{\mu\nu\rho\sigma}^{'abcd},
\label{Xprime}
\end{align}
where we have set $a= \frac{2b}{3} + c$, and $s$ represents a freely adjustable parameter that controls the weights with which the tensors  $E$ and $E'$ enters the definition of the vector $X'$. Then, $\Gamma_{\mu\nu\rho\sigma}^{abcd}(0,0,0,0)$ may be decomposed in terms 
of the color and Lorentz vectors $\Gz$, $\G$ and $X_{\mu\nu\rho\sigma}^{'abcd}$ as follows 
\be
\Gamma_{\mu\nu\rho\sigma}^{abcd}(0,0,0,0) = \frac{a-cs}2\Gz+
\left({\widetilde a}+\frac a2 + \frac{\widetilde b}{3}-\frac{cs}2\right) \G +c X_{\mu\nu\rho\sigma}^{'abcd}.
\ee 
Evidently, within the one-loop dressed approximation one has $c=0$.

\subsection{The case $(p,p,p,-3p)$}

We next turn to the case $(p_1,p_2,p_3,p_4)=(p,p,p,-3p)$.
In this case, the general form of the $V_i$ given \1eq{theV} remains the same, but now the  form factors are 
functions of the only available momentum scale, namely $p^2$, so that  $a_{ij} \to a_{ij}(p^2)$, $b_{ij} \to b_{ij}(p^2)$ and $c_{ij} \to c_{ij}(p^2)$.

In general, the presence of momenta makes the implementation of Bose symmetry more complicated, because it 
involves additional permutations of  $(p_1,p_2,p_3,p_4)$. However, for the particular case at hand, 
the fact that the form factors  can only depend on $p^2$, makes these momentum permutations ``inert''. 
As a result, one arrives at exactly the same form for $\Gamma_{\mu\nu\rho\sigma}^{abcd}(p,p,p,-3p)$ as the one given in  
\1eq{Gwith3}, with all coefficients converted into functions of $p^2$.

To see how this statement emerges from a more complete analysis, 
let us then focus, as before, on the $V_i^{\chic A}$ terms. 
Carrying out the same set of permutations as in the previous case, one may cast the  $V_i^{\chic A}$ in the form
\bea
(V_1^{\chic A})^{abcd}_{\mu\nu\rho\sigma} &=& 
g_{\mu\nu}g_{\rho\sigma} [{\widehat a}(p_2, p_3, p_1, p_4) \delta^{ab}\delta^{cd} + {\widetilde a}(p_1, p_3, p_2, p_4) \delta^{ac}\delta^{bd} 
+ {\widetilde a}(p_2, p_3, p_1, p_4)\delta^{ad}\delta^{bc} ],
\nonumber\\
(V_2^{\chic A})^{abcd}_{\mu\nu\rho\sigma} &=& 
g_{\mu\rho}g_{\nu\sigma} [ {\widetilde a} (p_1, p_2, p_3, p_4)\delta^{ab}\delta^{cd} + {\widehat a}(p_1, p_2, p_3, p_4)\delta^{ac}\delta^{bd} + 
{\widetilde a}(p_3, p_2, p_1, p_4)\delta^{ad}\delta^{bc}],
\nonumber\\
(V_3^{\chic A})^{abcd}_{\mu\nu\rho\sigma} &=& 
g_{\mu\sigma}g_{\nu\rho} [ {\widetilde a}(p_2, p_1, p_3, p_4)\delta^{ab}\delta^{cd} + {\widetilde a}(p_3, p_2, p_1, p_4)
\delta^{ac}\delta^{bd} + {\widehat a}(p_2, p_1, p_3, p_4)\delta^{ad}\delta^{bc}].
{}\nonumber\\
\eea
At this point it is clear that if we choose the kinematics $p_1= p_2=p_3=p$ and $p_4=-3p$, and since $p_4$ appears always last 
in all the arguments of the form factors, after setting 
${\widehat a}(p, p, p, -3p) \equiv {\widehat a}(p^2)$ and ${\widetilde a}(p, p, p, -3p) \equiv {\widetilde a}(p^2)$, 
the above expressions reduce to 
\bea
(V_1^{\chic A})^{abcd}_{\mu\nu\rho\sigma} &=& 
g_{\mu\nu}g_{\rho\sigma} [{\widehat a}(p^2) \delta^{ab}\delta^{cd} + {\widetilde a}(p^2) \delta^{ac}\delta^{bd} 
+ {\widetilde a}(p^2)\delta^{ad}\delta^{bc} ],
\nonumber\\
(V_2^{\chic A})^{abcd}_{\mu\nu\rho\sigma} &=& 
g_{\mu\rho}g_{\nu\sigma} [ {\widetilde a} (p^2)\delta^{ab}\delta^{cd} + {\widehat a}(p^2)\delta^{ac}\delta^{bd} + 
{\widetilde a}(p^2)\delta^{ad}\delta^{bc}],
\nonumber\\
(V_3^{\chic A})^{abcd}_{\mu\nu\rho\sigma} &=& 
g_{\mu\sigma}g_{\nu\rho} [ {\widetilde a}(p^2)\delta^{ab}\delta^{cd} + {\widetilde a}(p^2)
\delta^{ac}\delta^{bd} + {\widehat a}(p^2)\delta^{ad}\delta^{bc}].
{}\nonumber\\
\eea
Thus, after the definition $a(p^2) \equiv  {\widehat a}(p^2) -  {\widetilde a}(p^2)$, 
one arrives at exactly the same expression as in \1eq{V1con},
with the only difference that the coefficients are now functions of $p^2$. 
The same conclusions are reached for the $V_i^{\chic B}$, where $b\to b(p^2)$ and ${\widetilde b}\to {\widetilde b}(p^2)$, 
while $V_i^{\chic C}$ vanishes as before.

At this point it may seem that the above construction hinges on 
the fact that the permutations chosen are such that $p_4$ appears always last. This is, however, not so;
indeed, one may carry out all 24 possible permutations (which inevitably place  $p_4$ in all possible positions),  
imposing every time the requirement of Bose symmetry, arriving at the following exhaustive set of conditions:

\bea
{\widetilde a}(p_1, p_2, p_3, p_4) &=& {\widetilde a}(p_3, p_2, p_1, p_4) = {\widetilde a}(p_2, p_1, p_4, p_3)= {\widetilde a}(p_4, p_1, p_2, p_3)=
\nonumber\\
{\widetilde a}(p_1, p_4, p_3, p_2) &=& {\widetilde a}(p_3, p_4, p_1, p_2) ={\widetilde a}(p_2, p_3, p_4, p_1)={\widetilde a}(p_4, p_3, p_2, p_1),  
\nonumber\\
{\widetilde a}(p_2, p_3, p_1, p_4) &=& {\widetilde a}(p_1, p_3, p_2, p_4) = {\widetilde a}(p_2, p_4, p_1, p_3)= {\widetilde a}(p_1, p_4, p_2, p_3)=
\nonumber\\
{\widetilde a}(p_4, p_1, p_3, p_2) &=& {\widetilde a}(p_3, p_1, p_4, p_2) ={\widetilde a}(p_4, p_2, p_3, p_1)={\widetilde a}(p_3, p_2, p_4, p_1),  
\nonumber\\
{\widetilde a}(p_2, p_1, p_3, p_4) &=& {\widetilde a}(p_3, p_1, p_2, p_4) = {\widetilde a}(p_1, p_2, p_4, p_3)= {\widetilde a}(p_4, p_2, p_1, p_3)=
\nonumber\\
{\widetilde a}(p_4, p_3, p_1, p_2) &=& {\widetilde a}(p_1, p_3, p_4, p_2) ={\widetilde a}(p_2, p_4, p_3, p_1)={\widetilde a}(p_3, p_4, p_2, p_1),  
\label{permA}
\eea
and
\bea
{\widehat a}(p_1, p_2, p_3, p_4) &=& {\widehat a}(p_2, p_1, p_4, p_3) ={\widehat a}(p_3, p_4, p_1, p_2) = {\widehat a}(p_4, p_3, p_2, p_1) ,
\nonumber\\
{\widehat a}(p_1, p_3, p_2, p_4) &=& {\widehat a}(p_2, p_4, p_1, p_3) = {\widehat a}(p_3, p_1, p_4, p_2)= {\widehat a}(p_4, p_2, p_3, p_1) , 
\nonumber\\
{\widehat a}(p_2, p_3, p_1, p_4) &=& {\widehat a}(p_1, p_4, p_2, p_3) = {\widehat a}(p_4, p_1, p_3, p_2) = {\widehat a}(p_3, p_2, p_4, p_1) ,    
\nonumber\\
{\widehat a}(p_3, p_2, p_1, p_4) &=& {\widehat a}(p_4, p_1, p_2, p_3) ={\widehat a}(p_1, p_4, p_3, p_2) = {\widehat a}(p_2, p_3, p_4, p_1),
\nonumber\\
{\widehat a}(p_2, p_1, p_3, p_4) &=& {\widehat a}(p_1, p_2, p_4, p_3) ={\widehat a}(p_4, p_3, p_1, p_2) = {\widehat a}(p_3, p_4, p_2, p_1),
\nonumber\\
{\widehat a}(p_3, p_1, p_2, p_4) &=& {\widehat a}(p_4, p_2, p_1, p_3) ={\widehat a}(p_1, p_3, p_4, p_2) = {\widehat a}(p_2, p_4, p_3, p_1).
\label{permB}
\eea

Evidently, for the choice $p_1= p_2=p_3=p$ and $p_4=-3p$, each set of conditions listed in \1eq{permA} reduces, as anticipated, to  
the statement of complete equality, 
\be
{\widetilde a}(p, p, p, -3p) = {\widetilde a}(p, p, -3p, p)= {\widetilde a}(p, -3p, p, p)= {\widetilde a}(-3p, p, p, p),
\ee
and similarly from \1eq{permB}
\be
{\widehat a}(p, p, p, -3p) = {\widehat a}(p, p, -3p, p)= {\widehat a}(p, -3p, p, p)= {\widehat a}(-3p, p, p, p).
\ee

\section{\label{sif}Scalar integral functions}

In this Appendix we report the closed expressions for the various functions appearing in the calculations of Sect.~\ref{sec:nonzeroext}.

\subsection{Gluon boxes}

For the gluon boxes one has
\begin{align}
f_1(k,p)&=-8\frac{[(k\cd p)^2-k^2 p^2]^2}{(d-1)p^2}\{[6 (6 d-1) p^2+38 k^2] (k\cd p)^2+[(3 d+7) k^4+6 (5 d-2) k^2 p^2\nonumber \\
&+27 (d-1) p^4]p^2 +8[(3 d+5) k^2 p^2+9 (d-1) p^4+k^4] (k\cd p)+24 (k\cd p)^3\},\nonumber \\
f_2(k,p)&=12\frac{[(k\cd p)^2-k^2 p^2]^2}{(d^2-1)p^4}\{3 (d-1) k^8+6 k^6 [6 (d-1) (k\cd p)+(9 d-8) p^2]\nonumber \\
&+k^4[76 (5 d-4) p^2 (k\cd p)+6 (22 d-19)
   (k\cd p)^2+(247d-6 d^2-152) p^4]\nonumber \\
&+4 k^2 [(202d-12 d^2-47) p^4 (k\cd p)+(155 d-67) p^2 (k\cd p)^2+9 (4 d-1) (k\cd p)^3\nonumber \\
   &+3 (29d-5 d^2+7) p^6]+3[24 (7d-2 d^2+9) p^6 (k\cd p)+4 (31d-6 d^2+46) p^4 (k\cd p)^2\nonumber \\
   &+4 (8 d+29) p^2 (k\cd p)^3-18 (d-5) (d+1) p^8+45 (k\cd p)^4]
   \}.
\label{app:glbox}   
\end{align}

\subsection{Triangle diagrams}

For the class $(c^{\s{A}})$ graphs one has
\begin{align}
g_1(k,p)&=-2\frac{(k\cd p)^2-k^2 p^2}{(d-1)p^2}
\{-(d-2) k^6+k^4 [2 (d-3) p^2-6 (d-2) (k\cd p)]\nonumber \\
&+k^2 [4 (4 d-9) p^2 (k\cd p)+(18-8 d) (k\cd
   p)^2+3 (7 d-13) p^4]+(k\cd p) [(4 d-17) p^2 (k\cd p)\nonumber \\
   &+2 (d-8) p^4+4 (k\cd p)^2]\},\nonumber \\
g_2(k,p)&=-6\frac{(k\cd p)^2-k^2 p^2}{(d^2-1)p^4}
\{(-d^2+d-1) k^6 p^2+k^2 [2 \left(-4 d^2+2 d+3\right) p^4 (k\cd p)\nonumber \\
&+(-8 d^2+10 d+15)
   p^2 (k\cd p)^2-3 (d^2-1) p^6+12 (k\cd p)^3]+(k\cd p) [(4 d^2-d-5) p^4
   (k\cd p)\nonumber \\
   &+2 (d^2-d-2) p^6+2 (2 d+5) p^2 (k\cd p)^2+9 (k\cd p)^3]+k^4 [-6d (d-1) 
   p^2 (k\cd p)\nonumber \\
   &+2d (1-2 d)  p^4+3 (k\cd p)^2]\}.
\label{app:gltri-ca}      
\end{align}

The class $(c^{\s{B}})$ diagrams yields instead
\begin{align}
g'_1(k,p)&=2\frac{(k\cd p)^2-k^2 p^2}{(d-1)p^2}\{(d-2) k^6+2 k^4 [(d+16) (k\cd p)+(13 d-15) p^2]\nonumber \\
&+k^2 [4 (17 d-21) p^2 (k\cd p)+(54-8 d) (k\cd
   p)^2+(79 d-121) p^4]\nonumber \\
   &+(k\cd p) [(5-12 d) p^2 (k\cd p)+6 (d+12) p^4-88 (k\cd p)^2]\},
\nonumber \\
g'_2(k,p)&=6\frac{(k\cd p)^2-k^2 p^2}{(d^2-1)p^4}\{
k^2 p^2 [(d^2-d+1) k^4+2 (4 d^2+d-6) k^2 p^2+7 (d^2-1)p^4]\nonumber \\
&+[(-8 d^2-2 d+9) k^2 p^2
+(-12 d^2+5 d+17) p^4-3 k^4] (k\cd
   p)^2+2 p^2 [(d^2-d-2) k^4\nonumber\\
   &+(-2 d^2-2 d+3) k^2 p^2+3 (d^2-d-2)
   p^4] (k\cd p)+2 (4 d+1) p^2 (k\cd p)^3+3 (k\cd p)^4\}.
\label{app:gltri-cb}  
\end{align}

\subsection{Fishnet diagrams}

Finally, for the fishnet diagrams one gets
\begin{align}
h_1(k,p)&=\frac{1}{2(d-1)p^2}
\{(d^2+9 d-34) k^4 p^2+4 k^2 [(d^2+9 d-35) p^2 (k\cd p)\nonumber \\
&+(d^2+8 d-33) p^4
+6(k\cd p)^2]+4 (k\cd p)^2 [(d+23) p^2+25 (k\cd p)]\},
\nonumber \\
h_2(k,p)&=\frac{3}{2(d^2-1)p^4}\{
[4 \left(d^2-1\right) p^4-6 k^2 p^2] (k\cd p)^2+4 (d^3-2 d^2-2 d+1) k^2 p^4 (k\cd
   p)\nonumber \\
   &+k^2 p^4 [(d^3-2 d^2-d+5) k^2+4 (d^3-3 d^2-d+3) p^2]+4 (d+1) p^2 (k\cd
   p)^3+3 (k\cd p)^4.
\label{app:glfish}
\end{align}

It is straightforward but tedious to verify that, in the limit $p \to 0$, 
the above expressions reduce to the corresponding results found in Sect.~\ref{sec:zeroext}.

%\bibliography{../../../Bibliography/bibliography}

\end{document}